\documentclass[aps,pra,twocolumn,groupedaddress,amsmath,nofootinbib,showkeys]{revtex4-1}
\usepackage{graphicx}
\usepackage{subfigure}
\usepackage{mathtools}
\usepackage{mathrsfs} 
\usepackage{dcolumn}
\usepackage{bm}
\usepackage{amsmath}
\usepackage{color}
\usepackage{verbatim}
\usepackage{natbib} 
\usepackage{verbatim}
\usepackage{amsfonts}
\usepackage{lipsum}
\usepackage{epstopdf}
\usepackage[colorlinks,
            linkcolor=blue,
            anchorcolor=blue,
            citecolor=blue,
            urlcolor=blue]{hyperref}
\usepackage{bbm}

\begin{document}

\title{Quantum Geometric Tensor for Mixed States Based on the Covariant Derivative}

\author{Qianyi Wang}
\affiliation{National Laboratory of Solid State Microstructures, Key Laboratory of Intelligent Optical Sensing and Manipulation, 
College of Engineering and Applied Sciences and School of physics, and Collaborative Innovation Center of Advanced Microstructures, Nanjing University, Nanjing 210093, China}

\author{Ben Wang}
\email{ben.wang@nju.edu.cn}
\affiliation{National Laboratory of Solid State Microstructures, Key Laboratory of Intelligent Optical Sensing and Manipulation, 
College of Engineering and Applied Sciences and School of physics, and Collaborative Innovation Center of Advanced Microstructures, Nanjing University, Nanjing 210093, China}

\author{Jun Wang}
\affiliation{National Laboratory of Solid State Microstructures, Key Laboratory of Intelligent Optical Sensing and Manipulation, 
College of Engineering and Applied Sciences and School of physics, and Collaborative Innovation Center of Advanced Microstructures, Nanjing University, Nanjing 210093, China}

\author{Lijian Zhang}
\email{lijian.zhang@nju.edu.cn}
\affiliation{National Laboratory of Solid State Microstructures, Key Laboratory of Intelligent Optical Sensing and Manipulation, 
College of Engineering and Applied Sciences and School of physics, and Collaborative Innovation Center of Advanced Microstructures, Nanjing University, Nanjing 210093, China}
\date{\today}

\begin{abstract}
The quantum geometric tensor (QGT) is a fundamental quantity for characterizing the geometric properties of quantum states and plays an essential role in elucidating various physical phenomena. The traditional QGT, defined only for pure states, has limited applicability in realistic scenarios where mixed states are common. To address this limitation, we generalize the definition of the QGT to mixed states using the purification bundle and the covariant derivative. Notably, our proposed definition reduces to the traditional QGT when mixed states approach pure states. In our framework, the real and imaginary parts of this generalized QGT correspond to the Bures metric and the mean gauge curvature, respectively, endowing it with a broad range of potential applications. Additionally, using our proposed mixed-state QGT (MSQGT), we derive the geodesic equation applicable to mixed states. This work establishes a unified framework for the geometric analysis of both pure and mixed states, thereby deepening our understanding of the geometric properties of quantum states.

\end{abstract}

\keywords{quantum geometric tensor, covariant derivative, Bures metric}
\maketitle

\section{Introduction}

The quantum geometric tensor (QGT), a crucial concept connecting quantum mechanics and geometry, has achieved increasing attention recently \cite{QGT-Flaschner-Science,QGT-Wimmer-np,QGT-Yuyang-PRL,QGT-Gianfrate-nature,QGT-Yu-NSR,QGT-exp-2024}. The study of the QGT is based on the projective Hilbert space \cite{bengtsson2017geometry,nakahara2018geometry}, which is defined as the set of density operators for pure states. When this space is equipped with a positive-definite Riemannian metric and a symplectic form, it manifests as a Kähler manifold \cite{nakahara2018geometry}. The QGT combines these two geometric structures, thereby becoming a powerful tool for characterizing the geometric properties of quantum states \cite{QGT-introduction,QGT-introduction-review}. To date, the QGT has played a crucial role in understanding numerous physical phenomena, such as superfluidity in flat bands \cite{QGT-flat-bands-superfluidity-1,QGT-flat-bands-superfluidity-2}, orbital susceptibility \cite{QGT-Orbital-suscepitility-1,QGT-Orbital-suscepitility-2}, excitonic Lamb shift \cite{QGT-Lamb-shift}, the nonadiabatic anomalous Hall effect \cite{QGT-Orbital-suscepitility-1}, quantum phase transitions \cite{QGT-quantum-phase-transition}, quantum fluctuations \cite{QGT-quantum-fluctuation}, and topological matter \cite{QGT-topological-matter-1,QGT-topological-matter-2,QGT-topological-matter-3}.

Traditionally, the QGT has been defined solely for pure states. This limitation restricts its application in practical scenarios where mixed states—arising from decoherence or finite-temperature effects—are inevitable. To address this limitation, significant efforts have been made to develop meaningful definitions for a mixed-state quantum geometric tensor (MSQGT). For example, based on the Uhlmann bundle \cite{Uhlmann1986,Uhlmann1991,guo-hao-1}, a MSQGT was proposed with its real part corresponding to the well-known Bures metric \cite{guo-hao-1} or quantum Fisher information matrix (QFIM) \cite{Liu_JPA_2020}. This metric, related to fidelity \cite{Jozsa,Nielsen_Chuang_2010}, is widely used to quantify the distance between quantum states \cite{Uhlmann1976,Jozsa,Nielsen_Chuang_2010}. However, the imaginary part of this MSQGT invariably vanishes \cite{guo-hao-1}, contrasting with the pure-state QGT \cite{QGT-introduction,QGT-introduction-review}, whose imaginary part is the generally non-vanishing Berry curvature.
Based on the Sjöqvist distance \cite{sjoqvist-distance,guo-hao-2,guo-hao-3}, another type of MSQGT was introduced and can be regarded as the sum of the classical Fisher-Rao metric and the weighted pure-state QGT \cite{guo-hao-2}. Although this type of MSQGT can be reduced to the pure-state QGT, its real part does not equal the Bures metric, thereby limiting its applications.

In order to overcome these limitations, we introduce a new definition of MSQGT based on the covariant derivative. This definition not only maintains consistency with the QGT for pure states but also features a real part that equals the Bures metric, thereby offering a wide range of potential applications. Furthermore, leveraging the advantages of the covariant derivative, the geodesic equation for the space measured by the Bures metric can be readily derived. Moreover, through a systematic investigation of the horizontal lift, we identify a new local gauge-invariant quantity with potential for experimental verification.

The rest of this paper is organized as follows. Sec. \ref{sec:def-of-Uhlmann-bundle} introduces the purification bundle, based on the purification of mixed states, and defines its connection. In Sec. \ref{sec:def-of-covariant-derivative-and-MSQGT}, we define the covariant derivative based on this connection. Utilizing the gauge transformation properties of the covariant derivative, we then provide the definition of our MSQGT and analyze the characteristics of its real and imaginary parts. A key advantage of our MSQGT, emphasized in Sec. \ref{sec:relation-mixed-state-qgt-pure-state-qgt}, is its reduction to the traditional pure-state QGT as mixed states approach pure states. Sec. \ref{sec:geodesic}, as an important application of the covariant derivative, introduces a method for determining geodesics in the space measured by the Bures metric. In Sec. \ref{sec:holonomy}, we investigate the holonomy of closed curves and introduce a novel gauge-invariant quantity. Finally, Sec. \ref{sec:conclusion} concludes this paper. Detailed derivations are provided in the Appendices.

\section{Definition of the purification bundle and Connection}
\label{sec:def-of-Uhlmann-bundle}
For any mixed state of a quantum system $S$ with dimension $N$, it is always possible to purify the state by incorporating an environment system $E$ with the same dimension \cite{Nielsen_Chuang_2010}. The purification of the mixed state $\rho_S$ is represented by a pure state $|\psi\rangle$ for the combined system $S+E$, such that
$\mathrm{Tr}_E\bigl(|\psi\rangle\langle\psi|\bigr) = \rho_S$.
As stated by the Hughston-Joza-Wootters (HJW) theorem \cite{preskill1998lecture}, any two purifications of a density matrix can be connected by a unitary transformation acting on $E$.

\begin{figure}
    \centering
    \includegraphics[width=0.9\linewidth]{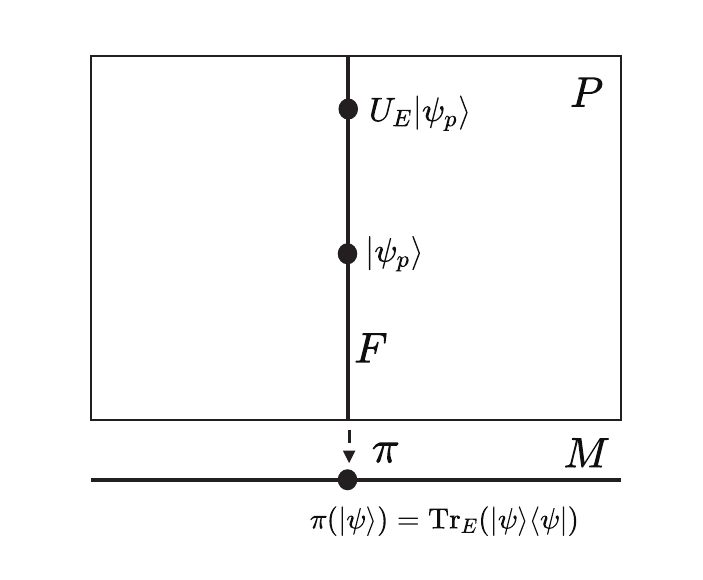}
    \caption{Illustration of the purification bundle. $P$ and $M$ represent the bundle manifold and base manifold respectively. $\pi$ is the projection map. $|\psi_p\rangle$ and $U_E|\psi_p\rangle$ belongs to the same fiber $F$ in the bundle manifold. $|\psi\rangle$ represents any point lies on the fiber $F$.}
    \label{fig:uhlmann-bundle}
\end{figure}

As depicted in FIG. \ref{fig:uhlmann-bundle}, the non-uniqueness of purification is naturally described within the framework of fiber bundle theory. Specifically, a bundle can be constructed by assigning the space of full-rank mixed states as the base manifold $M$, the space of purifications as the bundle manifold $P$, and the unitary group $\mathrm{U}(N)$ as the structure group. The projection map $\pi:P \to M$ is defined by the partial trace over the environment. Consequently, the fiber bundle constructed in this manner is termed the purification bundle.
In this bundle, each fiber comprises all pure states of the composite system $S+E$ that reduce to the same mixed state; different points within a fiber are connected by a unitary operator acting on $E$.

Another essential concept in fiber bundle theory is the connection, which dictates how fibers are glued together. In the purification bundle, where the bundle manifold comprises pure states, it is possible to determine the connection through an inner product. For any pair of tangent vectors $|X\rangle$ and $|Y\rangle$ in the tangent space of the point $|\psi_p\rangle$ in the bundle manifold, we can define a new inner product as follows:
\begin{equation}
    (|X\rangle, |Y\rangle) = \mathrm{Re} \langle X|Y \rangle,
    \label{eq:inner-product-def}
\end{equation}
where $\langle X|Y \rangle$ represents the standard Hilbert space inner product. We will establish the concept of horizontal vector based on this new inner product.

In the fiber bundle theory, a vertical curve is defined as a curve lies along a fiber. Thus, by the HJW theorem, any vertical curve through $|\psi_p\rangle$ in the purification bundle can be locally generated by a Hermitian operator, i.e.
\begin{equation}
    C_V(t)=I_S\otimes\exp(iH_E t)|\psi_p\rangle,
\end{equation}
where $H_E$ can be any Hermitian operator acting on $E$ and $t$ represents the curve's parameter. To simplify notation, we will omit $I_S$ in the remainder of this paper. Since the vertical vector is the tangent vector of a vertical curve, thus any vertical vector at $|\psi_p\rangle$ can be written as:
\begin{equation}
\frac{d}{dt}\bigg|_{t=0}C_V(t)=iH_E|\psi_p\rangle.
\label{eq:vertical-vector}
\end{equation}


By combining the vertical vectors and the new inner product defined in Eq. \eqref{eq:inner-product-def}, we can define the connection for the purification bundle. Specially, for a general curve $|\psi(t)\rangle$, we can define its connection $\mathcal{A}_t$ by
\begin{equation}
    i \mathcal{A}_t|\psi(t)\rangle=(|\partial_t\psi(t)\rangle)_V,
    \label{eq:vertical}
\end{equation}
where $|\partial_t\psi(t)\rangle$ denotes the tangent vector of curve $|\psi(t)\rangle$ and $(|\partial_t\psi(t)\rangle)_V$ is its vertical component. According to Eq. (\ref{eq:vertical-vector}), to ensure that $i\mathcal{A}_t|\psi(t)\rangle$ is vertical, the connection $\mathcal{A}_t$ must be a Hermitian operator on the environment.

Based on Eq. (\ref{eq:vertical}), we can derive the expression for the connection. Since the inner product of a horizontal tangent vector with a vertical one is zero, we obtain
\begin{equation}
    (iH_E|\psi\rangle,|\partial_t\psi\rangle)=(iH_E|\psi\rangle,i\mathcal{A}_t|\psi\rangle).
    \label{eq:connection-def}
\end{equation}
This identity must be satisfied for any Hermitian operator $H_E$ acting on the environment. Utilizing the Schmidt decomposition, the state $|\psi\rangle$ can be expressed as
\begin{equation}
    |\psi\rangle =\sum_{i=0}^{N-1} \sqrt{p_i}|\xi_i\rangle|v_i\rangle,
    \label{eq:schmidt-decomposition}
\end{equation}
where $\{|\xi_i\rangle\}$ and $\{|v_i\rangle\}$ denote the orthonormal bases for $S$ and $E$, respectively. Then any Hermitian operator acting on $E$ can be decomposed as a real linear combination of the following operators: $h_{ik} = |v_i\rangle\langle v_k|+|v_k\rangle\langle v_i|$,$g_{ik} = i|v_i\rangle\langle v_k| - i|v_k\rangle\langle v_i|$,
so that the set $\{h_{ik},g_{ik}\}$ forms a basis for the space of Hermitian operators acting on $E$. Substituting this basis into Eq. (\ref{eq:connection-def}) one obtains
\begin{equation}
\mathcal{A}_{t}
=-i \left( \sum_{i=0}^{N-1} |\partial_{t}v_{i}\rangle  \langle  v_{i}|  + \sum_{i,k=0}^{N-1}\frac{2\sqrt{ p_{i}p_{k} }}{p_{i}+p_{k}} \langle  \xi_{k}|\partial_{t}\xi_{i}\rangle|v_{i}\rangle  \langle  v_{k}|  \right).
\label{eq:connection}
\end{equation}

To analyze how the connection transforms under local gauge transformations, consider applying a local gauge transformation to a given curve as follows:
\begin{equation}
|\psi(t)\rangle\to|\psi'(t)\rangle=U_E(t)|\psi(t)\rangle,
\label{eq:local-gauge-transformation}
\end{equation}
where $U_E(t)$ is a unitary operator acting on $E$. The new curve $ |\psi'(t)\rangle $ has the Schmidt decomposition
\begin{equation}
|\psi'(t)\rangle=\sum_{i=0}^{N-1} \sqrt{p_i} |\xi_i\rangle U_E(t)|v_i\rangle.
\end{equation}
Inserting this result into Eq. (\ref{eq:connection}) provides the local gauge transformations rule for the connection as:
\begin{equation}
\mathcal{A}_t\to\mathcal{A'}_t=U_E(t)\mathcal{A}_t U^\dagger_E(t)-i\partial_tU_E(t)U_E^\dagger(t).
\label{eq:gauge-transform-of-connection}
\end{equation}

It is important to note that the expression for the connection in Eq. (\ref{eq:connection}) presents significant challenges for numerical calculations. These challenges primarily stem from the fact that the Schmidt decomposition of $|\psi\rangle$ includes an inherent arbitrariness: for any phase $\phi_i$, both $\{e^{i\phi_i}|\xi_i\rangle\}$ and $\{e^{-i\phi_i}|v_i\rangle\}$ remain valid Schmidt bases for $|\psi\rangle$. Consequently, the derivative of Schmidt basis in Eq. (\ref{eq:connection}) cannot be determined uniquely. To address this issue, we reformulate Eq. (\ref{eq:connection}) into an expression that does not depend on the Schmidt decomposition:
\begin{equation}
\mathcal{A}_{t}=-i \mathscr{L}_{(\mathrm{Tr}_{S}|\psi \rangle  \langle \psi|)}(\mathrm{Tr}_{S}(|\partial_{t}\psi \rangle  \langle  \psi|-|\psi \rangle \langle  \partial_{t}\psi|)),
\label{eq:connection-calculate}
\end{equation}
The operator function $\mathscr{L}$ in this equation is defined as
\begin{equation}
\mathscr{L}_{\sigma}(O)=\sum_{i,k} \frac{\langle i|O|k\rangle}{q_{i}+q_{k}} |i\rangle  \langle  k|,
\end{equation}
where $\sigma$ is any mixed state and $|i\rangle$ denotes an eigenstate of $\sigma$, with $q_i$ being the eigenvalue associated with $|i\rangle$. The equivalence between Eq. \eqref{eq:connection} and Eq. \eqref{eq:connection-calculate} is demonstrated in Appendix \ref{appedix:equivalence-expression-connection}.

Consider a curve $\rho(t)$ on the base manifold and a curve $|\psi(t)\rangle$ on the bundle manifold, where $t$ denotes the curve parameter. If, for each $t$, the state $|\psi(t)\rangle$ purifies $\rho(t)$, we say that $|\psi(t)\rangle$ is a lift of $\rho(t)$. Moreover, if the tangent vector of $|\psi(t)\rangle$ remains horizontal at every point along the curve, then $|\psi(t)\rangle$ is called a horizontal lift. According to Eq. \eqref{eq:vertical}, the condition for a curve to be a horizontal lift is that its connection vanishes. For a horizontal lift $|\psi(t)\rangle$, the following relation holds:
\begin{equation}
\begin{aligned}
\left( |\psi(t)\rangle, |\psi(t+dt)\rangle \right)
&= \left| \langle \psi(t) | \psi(t+dt) \rangle \right|\\
&= F(\rho(t),\rho(t+dt)),
\label{eq:parallel-fidelity}
\end{aligned}
\end{equation}
where $F(A,B) = \mathrm{Tr}\Bigl(\sqrt{\sqrt{A}\,B\,\sqrt{A}}\Bigr)$ represents the fidelity between states $A$ and $B$. A detailed derivation of Eq. \eqref{eq:parallel-fidelity} is provided in Appendix \ref{appendix:Characteristics-of-horizontal-lift}.

\section{Covariant Derivative and the definition of the mixed-state quantum geometric tensor}
\label{sec:def-of-covariant-derivative-and-MSQGT}
We define the covariant derivative $|D_t\psi\rangle$ as the horizontal component of a tangent vector. Based on Eq. (\ref{eq:vertical}), the covariant derivative can be expressed by
\begin{equation}
|D_t\psi\rangle=|\partial_t\psi\rangle-(|\partial_t\psi\rangle)_V=|\partial_t\psi\rangle-i\mathcal{A}_t|\psi\rangle,
\label{eq:covariant-derivative-mixed}
\end{equation}
Suppose the mixed quantum states on the base manifold are parameterized by a set of parameters $\vec{x}=(x^1,x^2,...,x^\mu,..)^\top$. Then we can obtain that the covariant derivative shares properties similar to those of the ordinary derivative, including:
\begin{equation}
D_{\mu}\langle \psi|\psi \rangle =\langle  D_{\mu}\psi|\psi \rangle  +\langle  \psi|D_\mu \psi \rangle  =0,
\end{equation}
\begin{equation}
\begin{aligned}
D_{\mu}D_{\nu}\langle  \psi|\psi \rangle  
&= \langle  D_\mu D_{\nu}\psi|\psi \rangle +\langle  \psi|D_{\mu}D_{\nu}\psi \rangle\\
&+\langle  D_{\mu}\psi|D_{\nu}\psi \rangle  +\langle D_{\nu}\psi|D_{\mu}\psi \rangle=0,
\end{aligned}
\label{eq:second_covariant_derivative}
\end{equation}
where $D_\mu=\frac{\partial}{\partial x^\mu}-i\mathcal{A}_\mu$ and the second-order covariant derivative is defined as
\begin{equation}
\begin{aligned}
|D_\mu D_{\nu}\psi \rangle
&=|\partial_{\mu}\partial_{\nu}\psi \rangle  -i\partial_{\mu}\mathcal{A}_{\nu}|\psi \rangle  -i\mathcal{A}_{\nu}|\partial_{\mu}\psi \rangle  \\
&\qquad -i\mathcal{A}_{\mu}|\partial_{\nu}\psi \rangle  -\mathcal{A}_{\mu}\mathcal{A}_{\nu}|\psi \rangle.
\end{aligned}
\end{equation}
Moreover, under the local gauge transformation in Eq. (\ref{eq:local-gauge-transformation}), the covariant derivative transforms as
\begin{equation}
|D_\mu\psi\rangle\to|D'_\mu\psi'\rangle =|\partial_\mu\psi'\rangle-i\mathcal{A'}_\mu|\psi'\rangle=U_E(\vec{x})|D_\mu\psi\rangle.
\label{eq:gauge-transform-of-covariant-derivative}
\end{equation}
Thus, the covariant derivative transforms in the same manner as $|\psi\rangle$, which makes it frequently used in constructing gauge-invariant quantity.

\begin{figure*}[t]
    \centering
    \includegraphics[width=\linewidth]{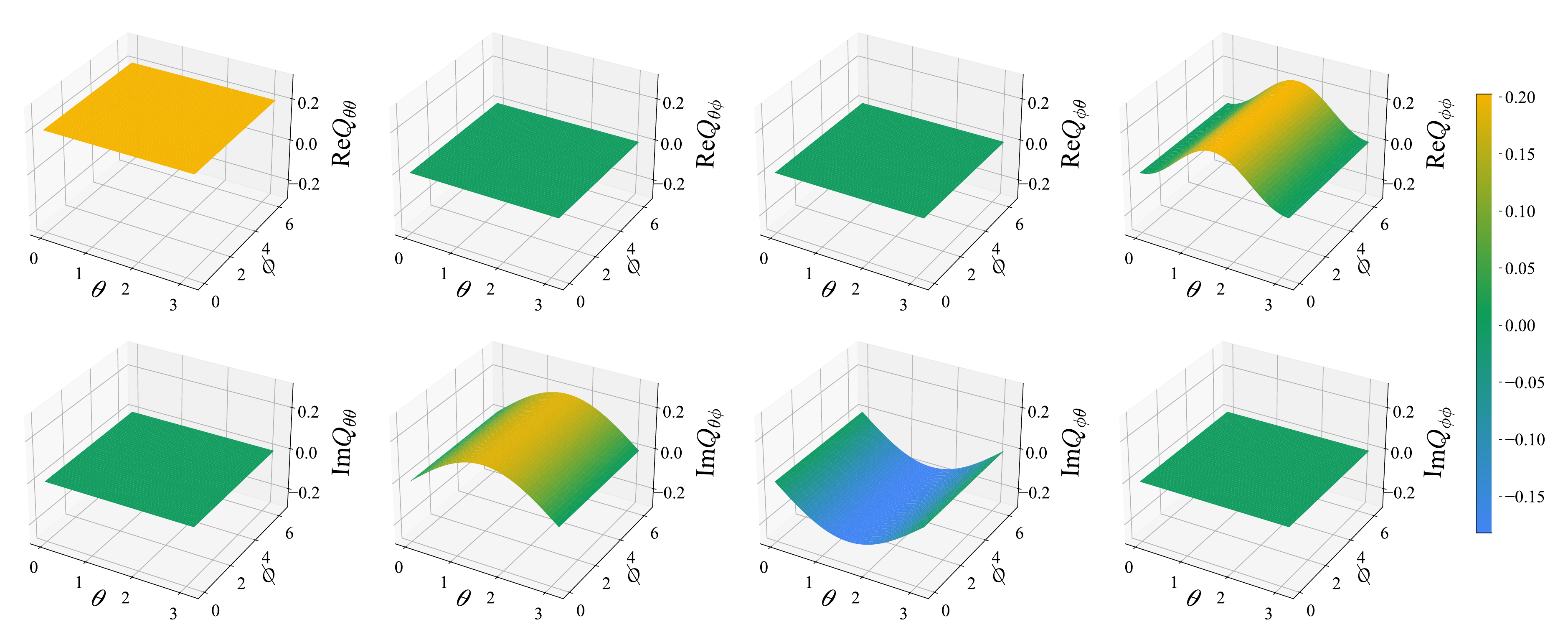}
    \caption{Quantum geometric tensor for mixed qubit states with a fixed purity equal to 0.95.}
    \label{fig:qgt}
\end{figure*}

Utilizing the covariant derivative, we can develop a gauge-invariant metric for the base manifold. The metric is defined as
\begin{equation}
Q_{\nu\mu}=\langle D_\nu\psi|D_\mu\psi\rangle.
\label{eq:QGT-def}
\end{equation}
This metric is precisely our new MSQGT and its gauge invariance can be readily examined using Eq. (\ref{eq:gauge-transform-of-covariant-derivative}). 
Substituting Eqs. (\ref{eq:schmidt-decomposition}) and (\ref{eq:connection}) into this definition yields:
\begin{equation}
\begin{aligned}
Q_{\nu \mu}&=\sum_{i=0}^{N-1}\partial_{\nu}\sqrt{ p_{i} }\partial_{\mu}\sqrt{ p_{i}} +\sum_{i=0}^{N-1}p_{i}\langle  \partial_{\nu}\xi_{i}|\partial_{\mu}\xi_{i}\rangle\\
&-\sum_{i,k=0}^{N-1} \frac{4p_{i}^{2}p_{k}}{(p_{i}+p_{k})^{2}} \langle  \partial_{\nu}\xi_{i}|\xi_{k}\rangle  \langle  \xi_{k}|\partial_{\mu}\xi_{i}\rangle.
\label{eq:QGT-expan}
\end{aligned}
\end{equation}
A detailed derivation of this expression can be found in Appendix \ref{appendix:derive-QGT}.

Denote the real and imaginary parts of our MSQGT by $g$ and $\sigma$ respectively, then $Q_{\nu\mu}=g_{\nu\mu}+i\sigma_{\mu\nu}$.
It can be inferred that $g_{\nu\mu}$ is symmetric, whereas the $\sigma_{\nu\mu}$ is antisymmetric under the exchange of indices $\nu$ and $\mu$, consistent with the pure-state QGT. The details of this derivation can be found in Appendix \ref{appendix:real-and-imaginary-part-derivation}. Considering the symmetry, the expression for the real part can be rewritten as
\begin{equation}
\begin{aligned}
g_{\nu \mu}&=\sum_{i=0}^{N-1} \partial_{\nu}\sqrt{ p_{i} }\partial_{\mu}\sqrt{ p_{i}}+\sum_{i=0}^{N-1}p_{i}\mathrm{Re}\langle  \partial_{\nu}\xi_{i}|\partial_{\mu}\xi_{i}\rangle \\
&\quad-\sum_{i,k=0}^{N-1} \frac{2p_{i}p_{k}}{p_{i}+p_{k}} \mathrm{Re}(\langle  \partial_{\nu}\xi_{i}|\xi_{k}\rangle  \langle  \xi_{k}|\partial_{\mu}\xi_{i}\rangle). 
\end{aligned}
\label{eq:qgt-real}
\end{equation}
This expression coincides with that for the Bures metric \cite{Liu-PA,Liu_JPA_2020}.

Similar to the pure-state QGT, the imaginary part of our MSQGT is closely related to the gauge curvature. The gauge curvature of the purification bundle is defined as the covariant exterior derivative of the connection:
\begin{equation}
\begin{aligned}
T_{\nu\mu}
=D_{\mu}\mathcal{A}_{\nu}-D_{\nu}\mathcal{A}_\mu
=\partial_{\mu}\mathcal{A}_{\nu}-\partial_{\nu}\mathcal{A}_{\mu}-i[\mathcal{A}_{\mu},\mathcal{A}_{\nu}].
\end{aligned}
\end{equation}
Since
$
{}[D_{\nu},D_{\mu}]|\psi \rangle  
=|D_{\nu}D_{\mu}\psi \rangle  -|D_{\mu}D_{\nu}\psi \rangle
=-i(\partial_{\nu}\mathcal{A}_{\mu}-\partial_{\mu}\mathcal{A}_{\nu})|\psi \rangle  -[\mathcal{A}_{\nu},\mathcal{A}_{\mu}]|\psi \rangle
$
holds for any purification $|\psi\rangle$, one can establish the following relation between the covariant derivative and the curvature:
\begin{equation}
i[D_{\nu},D_{\mu}]=\partial_{\nu}\mathcal{A}_{\mu}-\partial_{\mu}\mathcal{A}_{\nu}-i[\mathcal{A}_{\nu},\mathcal{A}_{\mu}]=T_{\nu\mu}.
\end{equation}
By combining this relation with Eqs. (\ref{eq:schmidt-decomposition}) and (\ref{eq:connection}), we can derive that
\begin{equation}
    \sigma_{\nu\mu}=\frac{1}{2}\langle \psi|i[D_\mu,D_\nu]|\psi\rangle=\frac{1}{2}\langle\psi|T_{\nu\mu}|\psi\rangle.
\end{equation}
The details of this derivation is provided in the supplementary material. Hence, the imaginary part of the MSQGT is equal to the mean gauge curvature.

Finally, it is noteworthy that while the expression for the MSQGT derived in Eq. \eqref{eq:QGT-expan} possesses theoretical significance, its practical utility for numerical computation is limited. This limitation originates from the inherent phase ambiguity in the eigenstates of a density matrix, which renders their derivatives non-unique. Therefore, directly calculating the MSQGT from Eq. \eqref{eq:QGT-expan} is not feasible, and an expression independent of this phase arbitrariness must be derived. Considering $\rho=\sum_i p_i|\xi_i\rangle\langle \xi_i|$, it can be easily shown that the MSQGT in Eq. (\ref{eq:QGT-expan}) is equal to the following expression:
\begin{equation}
Q_{\nu \mu}=\sum_{i,k=0}^{N-1} \frac{p_{i}}{(p_{i}+p_{k})^{2}} \langle  \xi_{i}|\partial_{\nu}\rho|\xi_{k}\rangle  \langle  \xi_{k}|\partial_{\mu}\rho|\xi_{i}\rangle.
\label{eq:QGT-calculate}
\end{equation}
The proof is provided in Appendix \ref{appendix:phase-arbitrariness-independent-qgt-appendix}. Clearly, this expression is unaffected by phase arbitrariness, giving it a wider range of applications. 

\section{Relation between Mixed-State and Pure-State Quantum Geometric Tensors}
\label{sec:relation-mixed-state-qgt-pure-state-qgt}
Although Eqs. \eqref{eq:QGT-expan} and \eqref{eq:QGT-calculate} were originally derived for full-rank density matrices, they can be asymptotically extended to pure-state cases. Specially, when $\rho$ approaches a pure state $|\xi_0\rangle\langle \xi_0|$, the MSQGT reduces to:
\begin{equation}
Q_{\nu \mu}\to\langle  \partial_{\nu}\xi_{0}|\partial_{\mu}\xi_{0}\rangle  -\langle\partial_{\nu}  \xi_{0}|\xi_{0}\rangle  \langle  \xi_{0}|\partial_{\mu}\xi_{0}\rangle,
\end{equation}
which precisely reproduces the traditional pure-state QGT \cite{QGT-introduction,QGT-introduction-review}. This agreement confirms the inherent consistency between our MSQGT and the pure-state QGT. A detailed derivation is provided in Appendix~\ref{appendix:mix-to-pure}.

To further investigate the relation, we explicitly compute the MSQGT for a qubit system under the following parametrization:
\begin{equation}
    \rho(\theta,\phi)=\frac{(I+\vec{r}\cdot \vec{\sigma})}{2},
\end{equation}
where $I$ is the identity, $\vec{\sigma} = (\sigma_x,\,\sigma_y,\,\sigma_z)$ denotes the vector composed of the Pauli matrices, and $\vec{r}=(r\sin\theta\cos\phi,r\sin\theta\sin\phi,r\cos\theta)$ is a three-dimensional vector called the Bloch vector.

We first analyze the case with $r$ fixed at $0.9$. Both the real and imaginary components of the MSQGT are numerically computed and presented in FIG. \ref{fig:qgt}. Clearly, the real components of $Q_{\theta\phi}$ displays symmetry ($\mathrm{Re}Q_{\theta\phi}=\mathrm{Re}Q_{\phi\theta}$) under the exchange of $\theta$ and $\phi$, while the imaginary component of $Q_{\theta\phi}$ is antisymmetric ($\mathrm{Im}Q_{\theta\phi}=-\mathrm{Im}Q_{\phi\theta}$). The diagonal elements of the imaginary part of the MSQGT vanishes ($\mathrm{Im}Q_{\theta\theta}=-\mathrm{Im}Q_{\phi\phi}=0$) due to the antisymmetry. Furthermore, the MSQGT exhibits $\phi$-independence under this parametrization.

\begin{figure}
    \centering
    \includegraphics[width=0.8\linewidth]{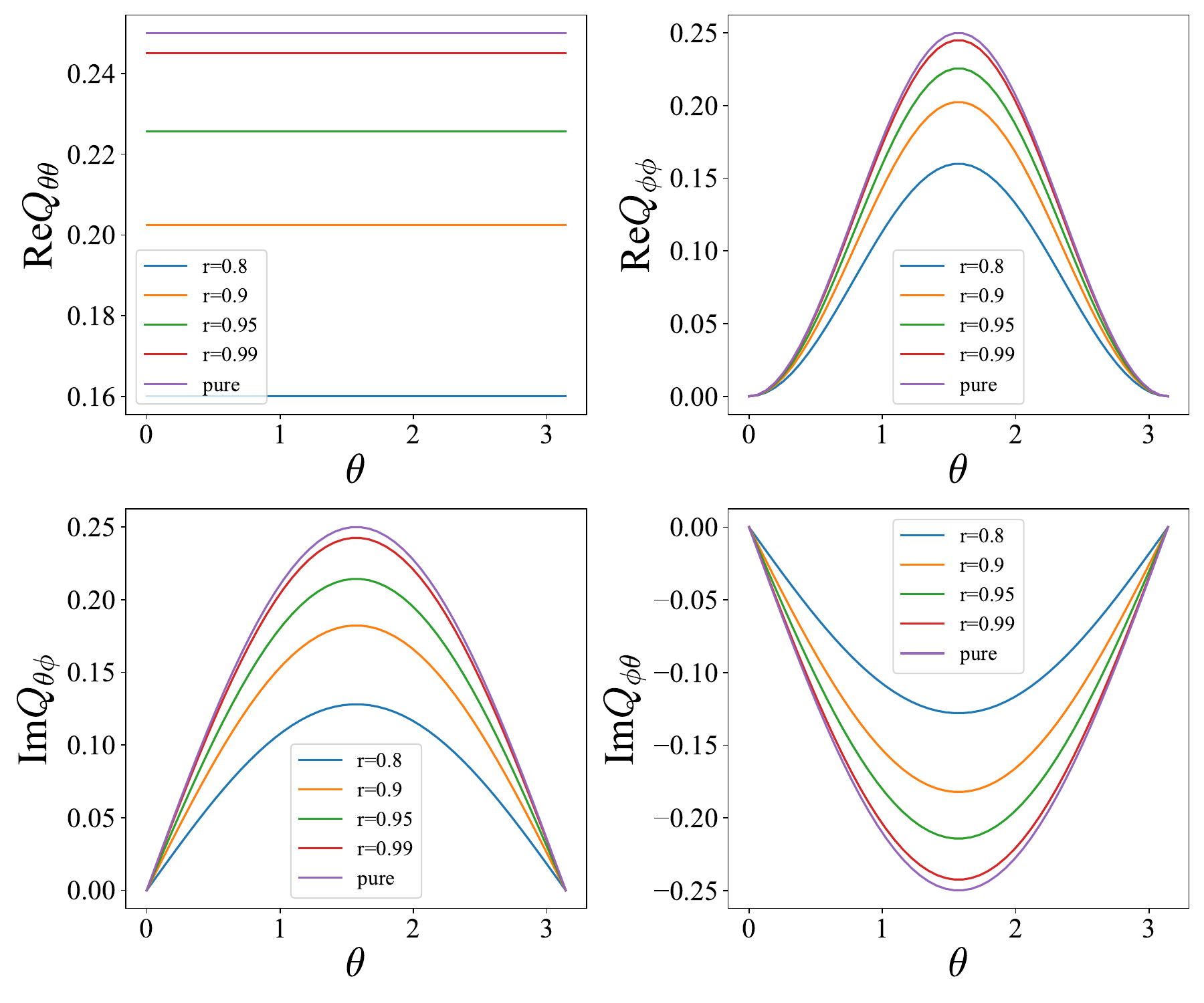}
    \caption{Comparison between the mixed-state quantum geometric tensor and pure-state quantum geometric tensor.}
    \label{fig:qgt-comparison}
\end{figure}

Finally, we compare the pure-state QGT and our MSQGT for various purities in FIG. \ref{fig:qgt-comparison} (Only the $\theta$-dependence is presented, as the QGT is independent of $\phi$, as indicated in FIG. \ref{fig:qgt}). Clearly, as the purity increases, the MSQGT approaches the traditional pure-state QGT. This confirms the theoretical prediction that the two definitions are consistent.

\section{Geodesics under the Metric of the MSQGT}
\label{sec:geodesic}

The real part of the newly defined MSQGT serves as the Bures metric on the base manifold. This naturally raises the question of how to determine geodesics with respect to the Bures metric. In this section, we derive the corresponding geodesic equation by exploiting the intrinsic properties of the covariant derivative and provide a general method to solve it.

Given an arbitrary curve $|\psi(t)\rangle$ on the bundle manifold with $t\in[0,T]$, its projection onto the base manifold has a length
\begin{equation}
l=\int_{0}^{T}\sqrt{g_{tt}dt^2}=\int_{0}^{T} \sqrt{\langle  D_{t}\psi|D_{t}\psi \rangle}  dt.
\label{eq:curve_length}
\end{equation}
Varying the curve length yields
\begin{equation}
\begin{aligned}
\delta l&=\int_{0}^{T} \frac{(\delta\langle   D_{t}\psi|)|D_{t}\psi \rangle +\langle  D_{t}\psi|(\delta |D_{t}\psi \rangle)  }{2\sqrt{ \langle  D_{t}\psi|D_{t}\psi \rangle}}dt\\
&=\int_{0}^{T} \frac{(|D_{t}\psi \rangle  ,\delta|D_{t}\psi \rangle  ) }{\sqrt{ \langle  D_{t}\psi|D_{t}\psi \rangle}}dt.
\end{aligned}
\end{equation}
Given that the length is invariant under reparametrization, $t$ can always be rescaled such that
\begin{equation}
\langle D_t\psi|D_t\psi\rangle = 1.
\label{eq:dl-rescale}
\end{equation}
Therefore,  it is sufficient to consider the variation
\begin{equation}
\delta l=\int_{0}^{T}(|D_{t}\psi \rangle  ,\delta|D_{t}\psi \rangle) dt.
\label{eq:variation of line length}
\end{equation}
To ensure that $\delta l=0$ for any $|\delta \psi\rangle$, $|\psi(t)\rangle$ must satisfy the condition:
\begin{equation}
|D_{t}D_{t}\psi \rangle=-|\psi\rangle.
\end{equation}
Details of this derivation are provided in Appendix \ref{appendix:geodesic}. This equation represents the geodesic equation under the Bures metric. It is important to note that this equation is valid only when the condition $\langle D_t\psi|D_t\psi\rangle = 1$ is satisfied. In the case where $|\psi\rangle$ is a horizontal lift with its connection vanishes, the geodesic equation can be further simplified to
\begin{equation}
|\partial_{t}\partial_{t}\psi \rangle=-|\psi \rangle,\quad  \text{(for horizontal lift of geodesics).}
\label{eq:geodesic}
\end{equation}
The general solution to this equation is given by
\begin{equation}
|\psi(t)\rangle=\cos(t)|\psi(0)\rangle+\sin(t)|\psi(\frac{\pi}{2})\rangle,
\label{eq:solution-of-geodesic}
\end{equation}
where $|\psi(0)\rangle$ and $|\psi(\pi/2)\rangle$ are constant quantum states whose determination requires appropriate boundary conditions.

Given two mixed states $\rho_A$ and $\rho_B$ separated by a Bures angle $\theta$ with $\theta<\pi/2$ (the geodesic connecting two points separated by $\pi/2$ is not unique), let $|\psi(t)\rangle$ denote the horizontal lift of the geodesic between them and $|\psi(t_A)\rangle$ and $|\psi(t_B)\rangle$ are purifications of $\rho_A$ and $\rho_B$ respectively. Without loss of generality, we can choose $t_A=0$. To determine $|\psi(t)\rangle$ completely, we need to determine $|\psi(\pi/2)\rangle$. Since the Bures angle between two points equals the path length of geodesic connecting them, according to Eqs. (\ref{eq:curve_length}) and (\ref{eq:dl-rescale}), we can obtain that 
\begin{equation}
\theta=l(\rho_A,\rho_B)=\int_0^{t_B}\sqrt{\langle D_t\psi(t)|D_t\psi(t)\rangle} dt=t_B.
\end{equation}
So $t_B=\theta$. Combining these conditions, $|\psi(\pi/2)\rangle$ can be determined by solving the following equations:
\begin{equation}
\begin{cases}
& \mathrm{Tr}_E(|\psi(\theta)\rangle\langle\psi(\theta)|)=\rho_B,\\
& \mathrm{Tr}_S(|\psi(0)\rangle\langle \psi(\frac{\pi}{2})|-|\psi(\frac{\pi}{2})\rangle\langle\psi(0)|)=0,\\
& \langle \psi(0)|\psi(\frac{\pi}{2})\rangle=0.\\
\end{cases}
\end{equation}
In these equations, the second equation ensures that $|\psi(t)\rangle$ is a horizontal lift and the third guarantees $\langle D_t \psi(t)|D_t \psi(t) \rangle=1$. After solving $|\psi(\pi/2)\rangle$, the geodesic connecting $\rho_A$ and $\rho_B$ can be obtained by projecting $|\psi(t)\rangle$ onto the base manifold.

As an example, we consider a qubit system and discuss the shape of geodesics in its Bloch sphere representation. 
One possible solution of the geodesic equation is 
\begin{equation}
\begin{aligned}
|\psi(t)\rangle&=\cos(t)\left(\sqrt{\frac{1+r}{2}}|00\rangle+\sqrt{\frac{1-r}{2}}|11\rangle\right)\\
&\qquad +\sin(t)\left(-\sqrt{\frac{1-r}{2}}|00\rangle+\sqrt{\frac{1+r}{2}}|11\rangle \right)
\end{aligned}
\end{equation}
where $|ij\rangle$ is the abbreviation of state $|i\rangle_S|j\rangle_E$ and $r$ can be chosen as any number between $0$ and $1$. Projecting $|\psi(t)\rangle$ onto the base manifold yields:
\begin{equation}
\rho(t)=\mathrm{Tr}_E(|\psi(t)\rangle\langle \psi(t)|)=\frac{I+\cos(2t+\varphi)\sigma_z}{2}
\end{equation}
with $\cos\varphi = r$. Clearly, the geodesic described by $\rho(t)$ corresponds to the $z$-axis. Considering the spherical symmetry of the Bloch sphere, we conclude that all diameters of the Bloch sphere are geodesics. 

Besides, another type of solution to the geodesic equation is
\begin{equation}
\begin{aligned}
|\psi'(t)\rangle&=\cos(t)\left(\sqrt{\frac{1+r}{2}}|00\rangle+\sqrt{\frac{1-r}{2}}|11\rangle\right)\\
&\qquad +\sin(t)\left(\sqrt{ \frac{1-r}{2} } |01\rangle  + \sqrt{ \frac{1+r}{2} }|10\rangle\right).
\end{aligned}
\end{equation}
By projecting it to the base manifold, we can obtain another geodesic
\begin{equation}
\rho'(t)=\frac{I+\sin(2t)\sigma_{x}+r \cos(2t)\sigma_{z}}{2},
\end{equation}
The geodesic traced by $\rho'(t)$ forms an ellipse described by the equation:
\begin{equation}
x^{2}+ \frac{z^2}{r^2}=1, \quad \forall r\in (0,1],
\end{equation}
where $x$ and $z$ denote the $x$- and $z$-components of the Bloch vector corresponding to $\rho'$.
Therefore, ellipses with a major axis length of one represent another category of geodesics in the Bloch sphere. It is important to note that diameters can also be considered as ellipses, characterized by a minor axis length of zero.
It is an interesting finding that geodesics under the Bures metric are typically ellipses rather than great circular arcs. Reduction to circular arcs occurs only when the two states involved are both pure.

\section{Holonomy in the purification bundle}
\label{sec:holonomy}
\begin{figure}[htbp]
    \centering
    \includegraphics[width=\linewidth]{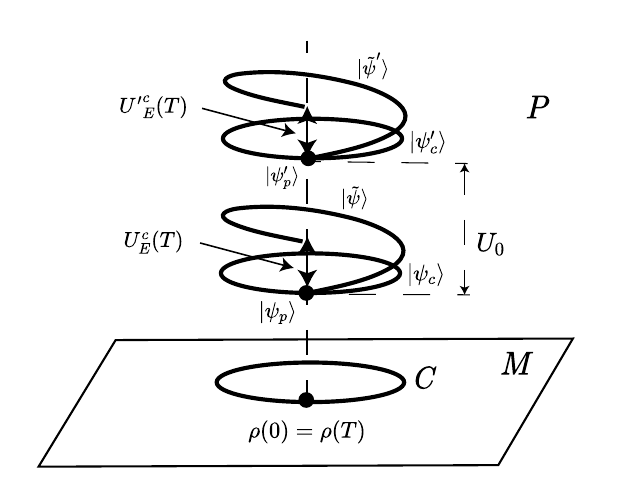}
    \caption{Illustration of the holonomy associated with the closed curve $C$. The curves $|\tilde{\psi}\rangle$ and $|\tilde{\psi}'\rangle$ denote two horizontally lifted curves passing through the states $|\psi_p\rangle$ and $|\psi'_p\rangle$, respectively. Correspondingly, the holonomies associated with these horizontal lifts are denoted by $U_E^c(T)$ and $U'^c_E(T)$. The curves $|\psi_c\rangle$ and $|\psi'_c\rangle$ describe two closed reference curves passing through $|\psi_p\rangle$ and $|\psi'_p\rangle$ respectively and they differ by a constant unitary transformation $U_0$.}
    \label{fig:holonomy}
\end{figure}

As shown in FIG. \ref{fig:holonomy}, we consider a closed curve $C$ on the base manifold. While the curve on the base manifold is closed, the starting and ending points of its horizontal lift typically do not coincide. Instead, they differ by a unitary operator acting on $E$, referred to as the holonomy. 

To obtain the expression for the holonomy, we select a closed reference path $|\psi_c(t)\rangle$ in the bundle manifold, with its projection being the closed curve $C$. Let us examine the horizontal lift of the curve $C$ originating from $|\psi_c(0)\rangle$. As depicted in FIG. \ref{fig:holonomy}, we denote this horizontal lift as $|\tilde{\psi}(t)\rangle$ and it satisfy $|\tilde{\psi}(0)\rangle=|\psi_c(0)\rangle$. The distinction between $|\psi_c(t)\rangle$ and $|\tilde{\psi}(t)\rangle$ is represented by $U_E^c(t)$ i.e.
\begin{equation}
|\tilde{\psi}(t)\rangle=U_E^c(t)|\psi_c(t)\rangle.
\label{eq:horizontal-lift}
\end{equation}
Here, $\mathcal{A}^c_t$ and $\mathcal{A}_t$ denote the connections related to curves $|\psi_c(t)\rangle$ and $|\tilde{\psi}(t)\rangle$, respectively. Referring to Eq. \eqref{eq:gauge-transform-of-connection}, it follows that
\begin{equation}
\mathcal{A}_t = U_E^c(t) \mathcal{A}_t^c U_E^c(t)^\dagger - i \partial_t U_E^c(t)U_E^c(t)^\dagger.
\end{equation}
Since $|\tilde{\psi}(t)\rangle$ is a horizontal lift, it implies that $\mathcal{A}_t = 0$. Therefore, we have
\begin{equation}
i \partial_t U_E^c(t) = U_E^c(t) \mathcal{A}^c_t.
\end{equation}
By solving this equation, we derive 
\begin{equation}
U^c_E(T) = \left[\mathcal{P}\exp\left(i\int_0^T\mathcal{A}^c_\tau\, d\tau\right)\right]^\dagger,
\label{eq:holonomy}
\end{equation}
where $\mathcal{P}$ represents the path ordering. Given that the reference curve is closed, we have $|\psi_c(0)\rangle=|\psi_c(T)\rangle$, where $T$ represents the parameter corresponding to the ending point. According to Eq. (\ref{eq:horizontal-lift}), we can obtain that
\begin{equation}
|\tilde{\psi}(T)\rangle=U_E^c(T)|\psi_c(T)\rangle=U_E^c(T)|\psi_c(0)\rangle=U_E^c(T)|\tilde{\psi}(0)\rangle.
\end{equation}
Clearly, $U^c_E(T)$ represents the transformation between the starting and ending points of the horizontal lift, thereby precisely characterizing the holonomy of the closed curve $C$. Importantly, it should be emphasized that the holonomy $U^c_E(T)$ does not depend on the choice of reference curve but is determined solely by the chosen horizontal lift. Since only one horizontal lift of a closed curve exists through a given point \cite{spivak1979comprehensive}, the holonomy $U^c_E(T)$ depends only on the starting point of the selected horizontal lift.

It is worth noting that the holonomy is not invariant under gauge transformations. To illustrate this, as depicted in FIG. \ref{fig:holonomy}, we choose an alternative reference closed curve given by $|\psi'_c(t)\rangle=U_0|\psi_c(t)\rangle$ in the bundle manifold to compute the holonomy, where $U_0$ is a $t$ independent unitary operator. Based on Eqs. \eqref{eq:gauge-transform-of-connection} and \eqref{eq:holonomy}, the holonomy for the horizontal lift passing through $|\psi'_c(0)\rangle$ is
\begin{equation}
\begin{aligned}
{U'}^{c}_E(T)
&=\left[\mathcal{P}\exp\left(i\int_0^T U_0\mathcal{A}^c_\tau U_0^\dagger\, d\tau\right)\right]^\dagger\\
&= U_0\left[\mathcal{P}\exp\left(i\int_0^T\mathcal{A}^c_\tau\, d\tau\right)\right]^\dagger U_0^\dagger\\
&=U_0U^{c}_E(T)U_0^\dagger
\end{aligned}
\end{equation}
The equation describes the gauge transformation law for the holonomy. It is noteworthy that in the $U(1)$ gauge case, where $U_0$ is a complex number rather than a matrix, the holonomy becomes gauge-invariant and is recognized as the Aharonov-Anandan (A-A) geometric phase \cite{A-A-phase-1}.

Although the holonomy is generally not gauge invariant, the mean holonomy (or overlap between the starting and ending points of a horizontal lift of a closed curve) is gauge-invariant. Its expression is 
\begin{equation}
O_C=\langle \tilde{\psi}(0)|\tilde{\psi}(T)\rangle=\langle \tilde{\psi}(0)|U_E^c(T)|\tilde{\psi}(0)\rangle
\end{equation}
Under the gauge transformation of $U_0$, this quantity becomes
\begin{equation}
O_C'=\langle \tilde{\psi}(0)|U_0^\dagger U_0{U}^c_E(T)U_0^\dagger U_0|\tilde{\psi}(0)\rangle=O_C
\end{equation}
So, the mean holonomy is indeed gauge-invariant. Its phase is known as the Uhlmann phase \cite{guohao-scipost}.

\section{Conclusion}
\label{sec:conclusion}
In this paper, we propose a new definition for the mixed-state quantum geometric tensor (MSQGT). Our MSQGT converges to the traditional pure-state QGT when mixed states approach pure states. Its real part corresponds to the well-known Bures metric, while its imaginary part equals the mean gauge curvature. Furthermore, we derive the geodesic equation under the Bures metric and provide some solutions for qubit systems. Finally, we investigate the holonomy in the purification bundle and obtained a new gauge-invariant quantity. Our approach, based on the covariant derivative on the purification bundle, is a powerful tool for studying the gauge-invariant quantity and local geometrical properties of quantum states. This research enhances the understanding of the relationship between geometry and quantum mechanics and holds potential applications in diverse fields such as quantum information processing, quantum metrology, and quantum control. Future research is expected to focus on experimentally measuring the MSQGT and the mean holonomy related to a closed curve. Additionally, investigating the connections between the MSQGT and other fundamental concepts in quantum theory is a topic of further interest.

\begin{acknowledgments}
This work was supported by the National Natural Science Foundation   of   China (Grants No. 12347104, No. U24A2017, No. 12461160276, No. 12175075), the National Key Research and Development Program of China (Grants No. 2023YFC2205802), the Natural Science Foundation of Jiangsu Province (Grants No. BK20243060, BK20233001), in part by the State Key Laboratory of Advanced Optical Communication Systems and Networks, China.
\end{acknowledgments}


\begin{thebibliography}{10}
\expandafter\ifx\csname url\endcsname\relax
  \def\url#1{{\tt #1}}\fi
\expandafter\ifx\csname urlprefix\endcsname\relax\def\urlprefix{URL }\fi
\providecommand{\eprint}[2][]{\url{#2}}

\bibitem{QGT-Flaschner-Science}
Fläschner N, Rem B~S, Tarnowski M, Vogel D, Lühmann D~S, Sengstock K and
  Weitenberg C 2016 {\em Science\/} {\bf 352} 1091--1094
  \urlprefix\url{https://www.science.org/doi/abs/10.1126/science.aad4568}

\bibitem{QGT-Wimmer-np}
Wimmer M 2017 {\em Nat. Phys.\/} {\bf 13} 545--550 ISSN 1745-2481
  \urlprefix\url{https://doi.org/10.1038/nphys4050}

\bibitem{QGT-Yuyang-PRL}
Tan X, Zhang D~W, Yang Z, Chu J, Zhu Y~Q, Li D, Yang X, Song S, Han Z, Li Z,
  Dong Y, Yu H~F, Yan H, Zhu S~L and Yu Y 2019 {\em Phys. Rev. Lett.\/} {\bf
  122}(21) 210401
  \urlprefix\url{https://link.aps.org/doi/10.1103/PhysRevLett.122.210401}

\bibitem{QGT-Gianfrate-nature}
Gianfrate A 2020 {\em Nature\/} {\bf 578} 381--385 ISSN 1476-4687
  \urlprefix\url{https://doi.org/10.1038/s41586-020-1989-2}

\bibitem{QGT-Yu-NSR}
Yu M, Yang P, Gong M, Cao Q, Lu Q, Liu H, Zhang S, Plenio M~B, Jelezko F, Ozawa
  T, Goldman N and Cai J 2019 {\em Natl. Sci. Rev.\/} {\bf 7} 254--260 ISSN
  2095-5138 \urlprefix\url{https://doi.org/10.1093/nsr/nwz193}

\bibitem{QGT-exp-2024}
Kang M 2025 {\em Nat. Phys.\/} {\bf 21} 110--117 ISSN 1745-2481
  \urlprefix\url{https://doi.org/10.1038/s41567-024-02678-8}

\bibitem{bengtsson2017geometry}
Bengtsson I and {\.Z}yczkowski K 2017 {\em Geometry of quantum states: an
  introduction to quantum entanglement\/} (Cambridge university press)

\bibitem{nakahara2018geometry}
Nakahara M 2018 {\em Geometry, topology and physics\/} (CRC press)

\bibitem{QGT-introduction}
Cheng R 2013 Quantum geometric tensor (fubini-study metric) in simple quantum
  system: A pedagogical introduction (\textit{Preprint} \eprint{1012.1337})
  \urlprefix\url{https://arxiv.org/abs/1012.1337}

\bibitem{QGT-introduction-review}
Kolodrubetz M, Sels D, Mehta P and Polkovnikov A 2017 {\em Phys. Rep.\/} {\bf
  697} 1--87 ISSN 0370-1573 geometry and non-adiabatic response in quantum and
  classical systems
  \urlprefix\url{https://www.sciencedirect.com/science/article/pii/S0370157317301989}

\bibitem{QGT-flat-bands-superfluidity-1}
Peotta S and T{\"o}rm{\"a} P 2015 {\em Nature Communications\/} {\bf 6} 8944
  ISSN 2041-1723 \urlprefix\url{https://doi.org/10.1038/ncomms9944}

\bibitem{QGT-flat-bands-superfluidity-2}
Julku A, Peotta S, Vanhala T~I, Kim D~H and T\"orm\"a P 2016 {\em Phys. Rev.
  Lett.\/} {\bf 117}(4) 045303
  \urlprefix\url{https://link.aps.org/doi/10.1103/PhysRevLett.117.045303}

\bibitem{QGT-Orbital-suscepitility-1}
Gao Y, Yang S~A and Niu Q 2014 {\em Phys. Rev. Lett.\/} {\bf 112}(16) 166601
  \urlprefix\url{https://link.aps.org/doi/10.1103/PhysRevLett.112.166601}

\bibitem{QGT-Orbital-suscepitility-2}
Pi\'echon F, Raoux A, Fuchs J~N and Montambaux G 2016 {\em Phys. Rev. B\/} {\bf
  94}(13) 134423
  \urlprefix\url{https://link.aps.org/doi/10.1103/PhysRevB.94.134423}

\bibitem{QGT-Lamb-shift}
Srivastava A and Imamo\ifmmode~\breve{g}\else \u{g}\fi{}lu A~m~c 2015 {\em
  Phys. Rev. Lett.\/} {\bf 115}(16) 166802
  \urlprefix\url{https://link.aps.org/doi/10.1103/PhysRevLett.115.166802}

\bibitem{QGT-quantum-phase-transition}
Zanardi P, Giorda P and Cozzini M 2007 {\em Phys. Rev. Lett.\/} {\bf 99}(10)
  100603 \urlprefix\url{https://link.aps.org/doi/10.1103/PhysRevLett.99.100603}

\bibitem{QGT-quantum-fluctuation}
Wei M, Wang L, Wang B, Xiang L, Xu F, Wang B and Wang J 2023 {\em Phys. Rev.
  Lett.\/} {\bf 130}(3) 036202
  \urlprefix\url{https://link.aps.org/doi/10.1103/PhysRevLett.130.036202}

\bibitem{QGT-topological-matter-1}
Ahn J, Guo G~Y and Nagaosa N 2020 {\em Phys. Rev. X\/} {\bf 10}(4) 041041
  \urlprefix\url{https://link.aps.org/doi/10.1103/PhysRevX.10.041041}

\bibitem{QGT-topological-matter-2}
Roy R 2014 {\em Phys. Rev. B\/} {\bf 90}(16) 165139
  \urlprefix\url{https://link.aps.org/doi/10.1103/PhysRevB.90.165139}

\bibitem{QGT-topological-matter-3}
Bauer D, Talkington S, Harper F, Andrews B and Roy R 2022 {\em Phys. Rev. B\/}
  {\bf 105}(4) 045144
  \urlprefix\url{https://link.aps.org/doi/10.1103/PhysRevB.105.045144}

\bibitem{Uhlmann1986}
Uhlmann A 1986 {\em Rep. Math. Phys.\/} {\bf 24} 229--240 ISSN 0034-4877
  \urlprefix\url{https://www.sciencedirect.com/science/article/pii/0034487786900558}

\bibitem{Uhlmann1991}
Uhlmann A 1991 {\em Letters in Mathematical Physics\/} {\bf 21} 229--236 ISSN
  1573-0530 \urlprefix\url{https://doi.org/10.1007/BF00420373}

\bibitem{guo-hao-1}
Hou X~Y, Zhou Z, Wang X, Guo H and Chien C~C 2024 {\em Phys. Rev. B\/} {\bf
  110}(3) 035144
  \urlprefix\url{https://link.aps.org/doi/10.1103/PhysRevB.110.035144}

\bibitem{Liu_JPA_2020}
Liu J, Yuan H, Lu X~M and Wang X 2019 {\em J. Phys. A: Math. Theor.\/} {\bf 53}
  023001 \urlprefix\url{https://dx.doi.org/10.1088/1751-8121/ab5d4d}

\bibitem{Jozsa}
Jozsa R 1994 {\em J. Mod. Opt.\/} {\bf 41} 2315--2323
  \urlprefix\url{https://doi.org/10.1080/09500349414552171}

\bibitem{Nielsen_Chuang_2010}
Nielsen M~A and Chuang I~L 2010 {\em Quantum Computation and Quantum
  Information: 10th Anniversary Edition\/} (Cambridge University Press)

\bibitem{Uhlmann1976}
Uhlmann A 1976 {\em Rep. Math. Phys.\/} {\bf 9} 273--279 ISSN 0034-4877
  \urlprefix\url{https://www.sciencedirect.com/science/article/pii/0034487776900604}

\bibitem{sjoqvist-distance}
Sj\"oqvist E 2020 {\em Phys. Rev. Res.\/} {\bf 2}(1) 013344
  \urlprefix\url{https://link.aps.org/doi/10.1103/PhysRevResearch.2.013344}

\bibitem{guo-hao-2}
Zhou Z, Hou X~Y, Wang X, Tang J~C, Guo H and Chien C~C 2024 {\em Phys. Rev.
  B\/} {\bf 110}(3) 035404
  \urlprefix\url{https://link.aps.org/doi/10.1103/PhysRevB.110.035404}

\bibitem{guo-hao-3}
Wang X, Hou X~Y, Tang J~C and Guo H 2024 Mathematical foundation of the
  u$^n(1)$ quantum geometric tensor (\textit{Preprint} \eprint{2410.11664})
  \urlprefix\url{https://arxiv.org/abs/2410.11664}

\bibitem{preskill1998lecture}
Preskill J 1998 {\em California institute of technology\/} {\bf 16} 1--8

\bibitem{Liu-PA}
Liu J, Xiong H~N, Song F and Wang X 2014 {\em Physica A: Stat. Mech. its
  Appl.\/} {\bf 410} 167--173 ISSN 0378-4371
  \urlprefix\url{https://www.sciencedirect.com/science/article/pii/S0378437114003926}

\bibitem{spivak1979comprehensive}
Spivak M 1979 {\em Inc., Berkeley\/} {\bf 2}

\bibitem{A-A-phase-1}
Aharonov Y and Anandan J 1987 {\em Phys. Rev. Lett.\/} {\bf 58}(16) 1593--1596
  \urlprefix\url{https://link.aps.org/doi/10.1103/PhysRevLett.58.1593}

\bibitem{guohao-scipost}
Wang X, Hou X~Y, Zhou Z, Guo H and Chien C~C 2023 {\em SciPost Phys. Core\/}
  {\bf 6} 024
  \urlprefix\url{https://scipost.org/10.21468/SciPostPhysCore.6.1.024}

\end{thebibliography}
\bibliographystyle{iopart-num}

\appendix

\section{Proof of Equivalence of the Two Expressions for the Connection}
\label{appedix:equivalence-expression-connection}
In this Appendix, we provide the proof demonstrating the equivalence between Eq. \eqref{eq:connection} and Eq. \eqref{eq:connection-calculate}.
Consider a curve $\rho(t)$ in the base manifold, with its spectral decomposition given by
\begin{equation}
\rho(t)=\sum_{i=0}^{N-1} p_i|\xi_i\rangle\langle \xi_i|.
\end{equation}
According to the Schmidt decomposition, the purification of $\rho(t)$ in total space is
\begin{equation}
|\psi(t)\rangle=\sum_{i=0}^{N-1} \sqrt{p}_i |\xi_i\rangle|v_i\rangle,
\end{equation}
where $|\xi_i\rangle$ and $|v_i\rangle$ are basis states of system and environment respectively. Then
\begin{equation}
\begin{aligned}
|\partial_{t}\psi \rangle  &=\sum_{i=0}^{N-1} \partial_{t}\sqrt{ p_{i} }|\xi_{i}\rangle  |v_{i}\rangle  +\sum_{i=0}^{N-1} \sqrt{ p_{i} }|\partial_{t}\xi_{i}\rangle  |v_{i}\rangle  \\
&\qquad\qquad +\sum_{i=0}^{N-1} \sqrt{ p_{i} }|\xi_{i}\rangle  |\partial_{t}v_{i}\rangle,
\end{aligned}
\end{equation}
\begin{equation}
\begin{aligned}
|\partial_{t}\psi \rangle  \langle  \psi|
&=\sum_{i,j=0}^{N-1} \sqrt{ p_{j} }\partial_{t}\sqrt{ p_{i} }|\xi_{i}\rangle  |v_{i}\rangle  \langle  \xi_{j}|\langle  v_{j}|\\
&\qquad+\sum_{i,j=0}^{N-1} \sqrt{ p_{i}p_{j} } |\partial_{t}\xi_{i}\rangle  |v_{i}\rangle  \langle  \xi_{j}|\langle v_{j}|\\
&\qquad + \sum_{i,j=0}^{N-1} \sqrt{ p_{i}p_{j} } |\xi_{i}\rangle  |\partial_{t}v_{i}\rangle  \langle  \xi_{j}|\langle  v_{j}|,
\end{aligned}
\end{equation}
\begin{equation}
\begin{aligned}
\mathrm{Tr}_{S}(|\partial_{t}\psi \rangle  \langle  \psi|)
&=\sum_{i=0}^{N-1} \sqrt{ p_{i} }\partial_{t}\sqrt{p_{i}}|v_{i}\rangle  \langle  v_{i}|\\
&\qquad+\sum_{i,j=0}^{N-1} \sqrt{ p_{i}p_{j} } \langle  \xi_{j}|\partial_{t}\xi_{i}\rangle  |v_{i}\rangle  \langle  v_{j}|\\
&\qquad+\sum_{i=0}^{N-1} p_{i} |\partial_{t}v_{i}\rangle  \langle  v_{i}|,
\end{aligned}
\end{equation}
\begin{equation}
\begin{aligned}
&\mathrm{Tr}_{S}(|\psi \rangle  \langle  \partial_{t}\psi|)\\
&=\mathrm{Tr}_{S}(|\partial_t\psi \rangle  \langle \psi|)^\dagger\\
&=\sum_{i=0}^{N-1} \sqrt{ p_{i} }\partial_{t}\sqrt{ p_{i} }|v_{i}\rangle  \langle  v_{i}|\\
&\qquad+\sum_{i,j=0}^{N-1} \sqrt{ p_{i}p_{j} } \langle  \partial_{t}\xi_{i}|\xi_{j}\rangle  |v_{j}\rangle  \langle  v_{i}|+\sum_{i=0}^{N-1} p_{i}|v_{i}\rangle  \langle  \partial_{t}v_{i}|\\
&=\sum_{i=0}^{N-1} \sqrt{ p_{i} }\partial_{t}\sqrt{ p_{i} }|v_{i}\rangle  \langle  v_{i}|\\
&\qquad-\sum_{i,j=0}^{N-1} \sqrt{ p_{i}p_{j} } \langle  \xi_{j}|\partial_{t}\xi_{i}\rangle  |v_{i}\rangle  \langle  v_{j}|+\sum_{i=0}^{N-1} p_{i}|v_{i}\rangle  \langle  \partial_{t}v_{i}|.
\label{eq:par-psi}
\end{aligned}
\end{equation}
In Eq. \eqref{eq:par-psi}, we have used the condition $\partial_t\langle\xi_i|\xi_j\rangle=\langle \partial_t \xi_i|\xi_j\rangle+\langle \xi_i|\partial_t\xi_j\rangle=\partial_t \delta_{ij}=0$. 
The difference between $\mathrm{Tr}_{S}(|\partial_{t}\psi \rangle  \langle\psi|)$ and $\mathrm{Tr}_{S}(|\psi \rangle  \langle  \partial_{t}\psi|)$ is
\begin{equation}
\begin{aligned}
&\mathrm{Tr}_{S}(|\partial_{t}\psi \rangle  \langle  \psi|-|\psi \rangle  \langle  \partial_{t}\psi|)\\
&=\sum_{i,j=0}^{N-1} 2\sqrt{ p_{i}p_{j} } \langle  \xi_{j}|\partial_{t}\xi_{i}\rangle|v_{i}\rangle  \langle  v_{j}| \\
&+\sum_{i=0}^{N-1}p_{i} (|\partial_{t} v_{i}\rangle  \langle  v_{i}|-|v_{i}\rangle  \langle  \partial_{t}v_{i}|).
\end{aligned}
\end{equation}
Denote $\sigma=\mathrm{Tr}_S(|\psi\rangle\langle \psi|)$, it can be easily calculated that
\begin{equation}
\sigma=\sum_{i=0}^{N-1} p_i|v_i\rangle\langle v_i|.
\end{equation}
Then
\begin{equation}
\begin{aligned}
&\mathscr{L}_{\sigma} \mathrm{Tr}_{S}(|\partial_{t}\psi \rangle  \langle  \psi|-|\psi \rangle  \langle  \partial_{t}\psi|)\\
&=\sum_{i,j,k,l=0}^{N-1} \frac{2\sqrt{ p_{i}p_{j} }}{p_{k}+p_{l}} \langle  \xi_{j}|\partial_{t}\xi_{i}\rangle  \langle  v_{k}|v_{i}\rangle  \langle  v_{j}|v_{l}\rangle  |v_{k}\rangle  \langle  v_{l}|\\
&\qquad+\sum_{i,k,l=0}^{N-1} \frac{p_{i}}{p_{k}+p_{l}}\langle  v_{k}|\partial_{t}v_{i}\rangle  \langle  v_{i}|v_{l}\rangle  |v_{k}\rangle  \langle  v_{l}|\\
&\qquad-\sum_{i,k,l=0}^{N-1} \frac{p_{i}}{p_{k}+p_{l}}\langle  v_{k}|v_{i}\rangle  \langle  \partial_{t}v_{i}|v_{l}\rangle  |v_{k}\rangle  \langle  v_{l}|\\
&=\sum_{i,j=0}^{N-1} \frac{2\sqrt{ p_{i}p_{j} }}{p_{i}+p_{j}} \langle  \xi_{j}|\partial_{t}\xi_{i}\rangle   |v_{i}\rangle  \langle  v_{j}|\\
&\qquad +\sum_{i,k=0}^{N-1} \frac{p_{i}}{p_{k}+p_{i}}\langle  v_{k}|\partial_{t}v_{i}\rangle   |v_{k}\rangle  \langle  v_{i}|\\
&\qquad-\sum_{i,l} \frac{p_{i}}{p_{i}+p_{l}}  \langle  \partial_{t}v_{i}|v_{l}\rangle  |v_{i}\rangle  \langle  v_{l}|)\\
&=\sum_{i,j=0}^{N-1} \frac{2\sqrt{ p_{i}p_{j} }}{p_{i}+p_{j}} \langle  \xi_{j}|\partial_{t}\xi_{i}\rangle   |v_{i}\rangle  \langle  v_{j}|\\
&\qquad+\sum_{i,k=0}^{N-1} \frac{p_{i}}{p_{k}+p_{i}}\langle  v_{k}|\partial_{t}v_{i}\rangle   |v_{k}\rangle  \langle  v_{i}|\\
&\qquad+\sum_{i,k=0}^{N-1} \frac{p_{k}}{p_{k}+p_{i}}  \langle  v_{k}|\partial_{t}v_{i}\rangle  |v_{k}\rangle  \langle  v_{i}|)\\
&=\sum_{i,j=0}^{N-1} \frac{2\sqrt{ p_{i}p_{j} }}{p_{i}+p_{j}} \langle  \xi_{j}|\partial_{t}\xi_{i}\rangle   |v_{i}\rangle  \langle  v_{j}|+\sum_{i,k=0}^{N-1} \langle  v_{k}|\partial_{t}v_{i}\rangle   |v_{k}\rangle  \langle  v_{i}|\\
&=\sum_{i,j=0}^{N-1} \frac{2\sqrt{ p_{i}p_{j} }}{p_{i}+p_{j}} \langle  \xi_{j}|\partial_{t}\xi_{i}\rangle   |v_{i}\rangle  \langle  v_{j}|+\sum_{i=0}^{N-1} |\partial_{t}v_{i}\rangle    \langle  v_{i}|,
\end{aligned}
\end{equation}
where we have used the completeness relation $\sum_{i=0}^{n-1}|v_k\rangle\langle v_k|=I_E$.
Clearly, $-i \mathscr{L}_{\sigma}(|\partial_t\psi\rangle\langle\psi|-|\psi\rangle\langle \partial_t\psi|)$ equals to the connection expressed in Eq. \eqref{eq:connection}.

\section{Characteristics of the Horizontal Lift}
\label{appendix:Characteristics-of-horizontal-lift}
Consider a curve $|\psi(t)\rangle$ in the bundle manifold, where $|\psi(t)\rangle$ represents the purification of $\rho(t)$. The condition for $|\psi(t)\rangle$ to be a horizontal lift is given by the vanishing of connection, i.e.,
\begin{equation}
\begin{aligned}
\mathcal{A}_{t}
&=-i \left( \sum_{i=0}^{N-1} |\partial_{t}v_{i}\rangle  \langle  v_{i}|  + \sum_{i,k=0}^{N-1}\frac{2\sqrt{ p_{i}p_{k} }}{p_{i}+p_{k}} \langle  \xi_{k}|\partial_{t}\xi_{i}\rangle|v_{i}\rangle  \langle  v_{k}|  \right)\\
&=0.
\label{eq:connection-appendix}
\end{aligned}
\end{equation}
We have assumed that the Schmidt decomposition of $|\psi(t)\rangle$ is
$|\psi(t)\rangle=\sum_i^{N-1} \sqrt{p}_i |\xi_i\rangle|v_i\rangle$,
where $|\xi_i\rangle$ and $|v_i\rangle$ are basis states for system and environment respectively. 
By setting \eqref{eq:connection-appendix} to zero, we obtain
\begin{equation}
\langle v_{i}|\mathcal{A}_t|v_{i}\rangle = -i\Bigl(\langle v_{i}|\partial_{t}v_{i}\rangle + \langle \xi_{i}|\partial_{t}\xi_{i}\rangle\Bigr) = 0.
\end{equation}
Furthermore, by employing the normalization condition for the quantum state, $\sum_{i=0}^{N-1}\left(\sqrt{p_{i}}\right)^2 = 1$, we arrive at
\begin{equation}
\frac{1}{2}\partial_{t}\sum_{i=0}^{N-1}\left(\sqrt{p_{i}}\right)^2 = \sum_{i=0}^{N-1}\sqrt{p_{i}}\,\partial_{t}\sqrt{p_{i}} = 0.
\end{equation}
Then, for $|\psi(t)\rangle$, we have
\begin{equation}
\begin{aligned}
&\langle \psi(t)|\partial_{t}\psi(t)\rangle\\
&=\Bigg(\sum_{j=0}^{N-1} \sqrt{p_j}\langle \xi_j|\langle v_j|\Bigg)\Bigg(\sum_{i=0}^{N-1}\partial_{t}\sqrt{ p_{i} }|\xi_{i}\rangle  |v_{i}\rangle \\
&\qquad\qquad+\sum_{i=0}^{N-1}\sqrt{ p_{i} }|\partial_{t}\xi_{i}\rangle  |v_{i}\rangle+\sum_{i=0}^{N-1}\sqrt{ p_{i} }|\xi_{i}\rangle  |\partial_{t}v_{i}\rangle\Bigg)  \\
&=\left(\sum_{i=0}^{N-1}\sqrt{p_{i}}\,\partial_{t}\sqrt{p_{i}} + \sum_{i=0}^{N-1}p_{i}\langle \xi_{i}|\partial_{t}\xi_{i}\rangle + \sum_{i=0}^{N-1}p_{i}\langle v_{i}|\partial_{t}v_{i}\rangle \right)\\
&=\left(\sum_{i=0}^{N-1}\sqrt{p_{i}}\,\partial_{t}\sqrt{p_{i}} + \sum_{i=0}^{N-1}p_{i}\Bigl(\langle \xi_{i}|\partial_{t}\xi_{i}\rangle + \langle v_{i}|\partial_{t}v_{i}\rangle\Bigr)\right)\\
&=0.
\end{aligned}
\end{equation}
Using this equation, we can further obtain 
\begin{equation}
\begin{aligned}
&(|\psi(t)\rangle,|\psi(t+dt)\rangle)\\
&=\frac{1}{2}(\langle \psi(t)|\psi(t+dt)\rangle+\langle \psi(t+dt)|\psi(t)\rangle)\\
&\approx \frac{1}{2}\left( 1+dt\langle \psi(t)|\partial_t\psi(t)\rangle+\frac{1}{2}dt^2\langle \psi(t)|\partial_t^2\psi(t)\rangle+h.c.\right)\\
&=\frac{1}{2}\left(1+\frac{1}{2}dt^2\langle \psi(t)|\partial_t^2\psi(t)\rangle+1+\frac{1}{2}dt^2\langle \partial_t^2\psi(t)|\partial\psi(t)\rangle\right)\\
&=1-\frac{1}{2}\langle\partial_t\psi(t)|\partial_t\psi(t)\rangle dt^2.
\label{eq:parallel-inner}
\end{aligned}
\end{equation}
where we have used
\begin{equation}
\partial_t^2\langle\psi|\psi\rangle=\partial_t^21=\langle \partial_t^2\psi|\psi\rangle+\langle \psi|\partial_t^2\psi\rangle+2\langle\partial_t\psi|\partial_t\psi\rangle=0.
\end{equation}
On the other hand
\begin{equation}
\begin{aligned}
&|\langle  \psi(t)|\psi(t+dt)\rangle|\\
&\approx|\langle  \psi(t)|(|\psi(t)\rangle  +dt \partial_{t}|\psi(t)\rangle  +dt ^{2}\frac{1}{2} \partial_{t}^{2}|\psi(t)\rangle  )|\\
&=\sqrt{ \left( 1+dt ^{2} \frac{1}{2}\langle  \psi|\partial_{t}^{2}\psi \rangle   \right)\left( 1+dt ^{2} \frac{1}{2} \langle  \partial_{t}^{2}\psi|\psi \rangle \right) }\\
&\approx\sqrt{ 1+dt ^{2} \frac{1}{2}(\langle  \psi|\partial_{t}^{2}\psi \rangle  +\langle  \partial_{t}^{2}\psi|\psi \rangle  ) }\\
&=\sqrt{ 1-dt ^{2}\langle \partial_t \psi|\partial_{t}\psi \rangle }\\
&\approx 1-\frac{1}{2} \langle\partial_t\psi|\partial_t\psi\rangle dt^2.
\end{aligned}
\end{equation}
Therefore, $(|\psi(t)\rangle,|\psi(t+dt)\rangle)=|\langle \psi(t)|\psi(t+dt)\rangle|$. 

Since the covariant derivative is $|D_t\psi\rangle=|\partial_t\psi\rangle-i\mathcal{A}_t|\psi\rangle$ and the connection vanishes, it follows that $|D_t\psi\rangle=|\partial_t\psi\rangle$. According to Eq. \eqref{eq:parallel-inner}, we obtain
\begin{equation}
\begin{aligned}
(|\psi(t)\rangle,|\psi(t+dt)\rangle)
&=1-\frac{1}{2}\langle D_t\psi|D_t\psi\rangle dt^2\\
&=1-\frac{1}{2}Q_{tt}dt^2\\
&=1-\frac{1}{2}g_{tt}dt^2\\
&=1-\frac{1}{2}ds_B^2(\rho(t),\rho(t+dt)).
\label{eq:parallel-inner-2}
\end{aligned}
\end{equation}
Here, $Q$ represents the MSQGT defined in the main text and $g$ is its real part. As stated in the main text, $g$ equals the Bures metric. Since the imaginary part of $Q$ is anti-symmetric, it makes no contribution to $Q_{tt}$. $ds_B$ is the Bures distance between $\rho(t)$ and $\rho(t+dt)$ and it equals to $\sqrt{2-2F(\rho(t),\rho(t+dt))}$ \cite{Liu_JPA_2020}, where $F(A,B) = \mathrm{Tr}\Bigl(\sqrt{\sqrt{A}\,B\,\sqrt{A}}\Bigr)$ represents the fidelity between states $A$ and $B$. Therefore, we deduce
\begin{equation}
(\langle \psi(t),|\psi(t+dt)\rangle)=|\langle \psi(t)|\psi(t+dt)\rangle|=F(\rho(t),\rho(t+dt)).
\label{eq:horizontal-lift-in-purification-bundle}
\end{equation}
This equation is exactly the Eq. \eqref{eq:parallel-fidelity} in the main text.

\section{Derivation of the Expression for the MSQGT}
\label{appendix:derive-QGT}
Suppose the mixed quantum states $\rho$ on the base manifold are parameterized by a set of parameters $\vec{x}$, and their spectral decomposition is
\begin{equation}
\rho(\vec{x})=\sum_i p_i|\xi_i\rangle\langle \xi_i|.
\end{equation}
According to the Schmidt decomposition, the purification of $\rho(\vec{x})$ in the bundle manifold can be written as
\begin{equation}
|\psi(\vec{x})\rangle=\sum_i \sqrt{p}_i |\xi_i\rangle|v_i\rangle,
\end{equation}
where $|v_i\rangle$ is basis state of environment. According to Eq. \eqref{eq:connection}, the connection for $|\psi(\vec{x})\rangle$ is
\begin{equation}
\mathcal{A}_\mu=-i \left(\sum_{i} |\partial_{\mu}v_{i}\rangle  \langle  v_{i}|+\sum_{i,k}  \frac{  2\sqrt{ p_{i}p_{k} } }{ p_{i}+p_{k}} \langle\xi_{k}|\partial_{\mu}\xi_{i}\rangle  |v_{i}\rangle  \langle  v_{k}|\right).
\end{equation}
Since
\begin{equation}
\begin{aligned}
|\partial_{\mu}\psi \rangle  &=\sum_{i=0}^{N-1} \partial_{\mu}\sqrt{ p_{i} }|\xi_{i}\rangle  |v_{i}\rangle+\sum_{i=0}^{N-1} \sqrt{ p_{i} }|\partial_{\mu}\xi_{i}\rangle  |v_{i}\rangle  \\
&\qquad +\sum_{i=0}^{N-1} \sqrt{ p_{i} } |\xi_{i}\rangle  |\partial_{\mu} v_{i}\rangle,
\end{aligned}
\end{equation}
and 
\begin{equation}
\begin{aligned}
&i\mathcal{A}_{\mu}|\psi \rangle \\ 
&=\sum_{i,l=0}^{N-1} |\partial_{\mu}v_{i}\rangle  \langle  v_{i}|\sqrt{ p_{l} }|\xi_{l}\rangle  |v_{l}\rangle  \\
&\qquad +\sum_{i,k,l=0}^{N-1} \frac{2\sqrt{ p_{i}p_{k} }}{p_{i}+p_{k}} \langle  \xi_{k}|\partial_{\mu}\xi_{i}\rangle  |v_{i}\rangle  \langle  v_{k}|\sqrt{ p_{l} } |\xi_{l}\rangle  |v_{l}\rangle  \\
&=\sum_{i=0}^{N-1} \sqrt{ p_{i} }|\xi_{i}\rangle|\partial_{\mu}v_{i}\rangle +\sum_{i,k=0}^{N-1} \frac{2\sqrt{ p_{i} }p_{k}}{p_{i}+p_{k}} \langle  \xi_{k}|\partial_{\mu}\xi_{i}\rangle  |\xi_{k}\rangle|v_{i}\rangle.
\end{aligned}
\end{equation}
Then, the covariant derivative is 
\begin{equation}
\begin{aligned}
&|D_{\mu}\psi \rangle\\
&=|\partial_\mu\psi\rangle-i\mathcal{A}_\mu|\psi\rangle\\
&=\sum_{i=0}^{N-1}\partial_{\mu}\sqrt{ p_{i} }|\xi_{i}\rangle  |v_{i}\rangle  -\sum_{i,k=0}^{N-1} \frac{2 \sqrt{ p_{i} }p_{k}}{p_{i}+p_{k}} \langle  \xi_{k}|\partial_{\mu}\xi_{i}\rangle  |\xi_{k}\rangle  |v_{i}\rangle\\ 
&\qquad+\sum_{i=0}^{N-1}\sqrt{ p_{i} }|\partial_{\mu}\xi_{i}\rangle  |v_{i}\rangle.
\label{eq:d_mu_psi}
\end{aligned}
\end{equation}
Taking the conjugate of both sides, we can obtain 
\begin{equation}
\begin{aligned}
\langle  D_{\nu}\psi|&=\sum_{i=0}^{N-1}\partial_{\nu}\sqrt{ p_{j} }\langle  \xi_{j}|  \langle  v_{j}|  -\sum_{jl=0}^{N-1}  \frac{2\sqrt{ p_{j}}p_{l}}{p_{j}+p_{l}} \langle  \partial_{\nu}\xi_{j}|\xi_{l}\rangle  \langle  \xi_{l}|  \langle v_{j}| \\
&\qquad +\sum_{j=0}^{N-1}\sqrt{ p_{j} }\langle  \partial_{\nu}\xi_{j}|  \langle  v_{j}|.
\label{eq:d_nu_psi}
\end{aligned}
\end{equation}
By substituting Eq. \eqref{eq:d_mu_psi} and Eq. \eqref{eq:d_nu_psi} into Eq. \eqref{eq:QGT-def}, we can obtain
\begin{equation}
\begin{aligned}
&\langle  D_{\nu}\psi|D_{\mu}\psi \rangle\\
&=\sum_{i=0}^{N-1} \partial_{\nu}\sqrt{ p_{i} }\partial_{\mu}\sqrt{ p_{i} }-\sum_{i} \sqrt{ p_{i} }\partial_{\mu}\sqrt{ p_{i} } \langle  \partial_{\nu}\xi_{i}|\xi_{i}\rangle  \\
&\quad+\sum_{i=0}^{N-1} \sqrt{ p_{i} }\partial_{\mu}\sqrt{ p_{i} } \langle  \partial_{\nu}\xi_{i}|\xi_{i}\rangle -\sum_{i=0}^{N-1} \sqrt{ p_{i} }\partial_{\nu}\sqrt{ p_{i} } \langle  \xi_{i}|\partial_{\mu}\xi_{i}\rangle\\
&\quad+\sum_{i,k=0}^{N-1} \frac{4p_{i}p_{k}^2}{(p_{i}+p_{k})^{2}}\langle  \xi_{k}|\partial_{\mu}\xi_{i}\rangle  \langle  \partial_{\nu}\xi_{i}|\xi_{k}\rangle\\
&\quad-\sum_{i,k=0}^{N-1} \frac{2p_{i}p_{k}}{p_{i}+p_{k}}\langle  \xi_{k}|\partial_{\mu}\xi_{i}\rangle  \langle  \partial_{\nu}\xi_{i}|\xi_{k}\rangle \\
&\quad+ \sum_{i=0}^{N-1} \sqrt{ p_{i} }\partial_{\nu}\sqrt{ p_{i} } \langle  \xi_{i}|\partial_{\mu}\xi_{i}\rangle \\
&-\sum_{i,k=0}^{N-1} \frac{2p_{i} p_{k}}{p_{i}+p_{k}}\langle  \partial_{\nu}\xi_{i}|\xi_{k}\rangle  \langle  \xi_{k}|\partial_{\mu}\xi_{i}\rangle +\sum_{i=0}^{N-1} p_{i}\langle  \partial_{\nu}\xi_{i}|\partial_{\mu}\xi_{i}\rangle\\  
&=\sum_{i=0}^{N-1}\partial_{\nu}\sqrt{ p_{i} }\partial_{\mu}\sqrt{ p_{i}} +\sum_{i=0}^{N-1}p_{i}\langle  \partial_{\nu}\xi_{i}|\partial_{\mu}\xi_{i}\rangle\\
&\quad -\sum_{i,k=0}^{N-1} \frac{4p_{i}p_{k}}{p_{i}+p_{k}}\langle  \partial_{\nu}\xi_{i}|\xi_{k}\rangle\langle  \xi_{k}|\partial_{\mu}\xi_{i}\rangle\\
&\quad+\sum_{i,k=0}^{N-1} \frac{4p_{i}p_{k}^{2}}{(p_{i}+p_{k})^{2}} \langle  \partial_{\nu}\xi_{i}|\xi_{k}\rangle  \langle  \xi_{k}|\partial_{\mu}\xi_{i}\rangle\\
&=\sum_{i=0}^{N-1}\partial_{\nu}\sqrt{ p_{i} }\partial_{\mu}\sqrt{ p_{i}} +\sum_{i=0}^{N-1}p_{i}\langle  \partial_{\nu}\xi_{i}|\partial_{\mu}\xi_{i}\rangle\\
&\quad-\sum_{i,k=0}^{N-1} \frac{4p_{i}^{2}p_{k}}{(p_{i}+p_{k})^{2}} \langle  \partial_{\nu}\xi_{i}|\xi_{k}\rangle  \langle  \xi_{k}|\partial_{\mu}\xi_{i}\rangle.
\label{eq:expan-of-qgt-appendix}
\end{aligned}
\end{equation}
This equation is exactly Eq. \eqref{eq:QGT-expan}.

\section{Properties of the Real and Imaginary Parts of the MSQGT}
\label{appendix:real-and-imaginary-part-derivation}
Denote the terms in Eq. (\ref{eq:QGT-expan}) as
\begin{equation}
H_{\nu \mu;i}=\langle  \partial_{\nu}\xi_{i}|\partial_{\mu}\xi_i
\rangle,\quad
R_{\nu\mu;ik}=\langle \partial_{\nu}\xi_{i}|\xi_{k}\rangle\langle \xi_{k}|\partial_{\mu}\xi_{i}\rangle.
\end{equation}
Since $\langle \partial_\nu\xi_i|\partial_\mu\xi_i\rangle=\langle \partial_\mu\xi_i|\partial_\nu\xi_i\rangle^*$, thus we have
\begin{equation}
\mathrm{Re}(H_{\nu\mu;i}) = \mathrm{Re}(H_{\mu\nu;i}),\quad \mathrm{Im}(H_{\nu\mu;i}) = -\mathrm{Im}(H_{\mu\nu;i}).
\label{eq:H-exchange-mu-nu}
\end{equation}
Next, consider the behavior of $R_{\nu\mu;ik}$ under the exchange $i$ and $k$. Since $\partial_{\rho}\langle  \xi_{i}|\xi_{k}\rangle  =\langle  \partial_{\rho}\xi_{i}|\xi_{k}\rangle  +\langle  \xi_{i}|\partial_{\rho}\xi_{k}\rangle=\partial_{\rho}\delta_{i,k}=0$,
we deduce that $\langle  \partial_{\rho}\xi_{i}|\xi_{k}\rangle=-\langle  \xi_{i}|\partial_{\rho}\xi_{k}\rangle$,
which further implies
$
R_{\nu\mu;ki}=\langle  \partial_{\nu}\xi_{k}|\xi_{i}\rangle  \langle  \xi_{i}|\partial_{\mu}\xi_{k}\rangle=\langle  \xi_{k}|\partial_{\nu}\xi_{i}\rangle  \langle \partial_{\mu} \xi_{i}|\xi_{k}\rangle=(\langle  \partial_{\nu}\xi_{i}|\xi_{k}\rangle  \langle  \xi_{k}|\partial_{\mu}\xi_{i}\rangle  )^* =R_{\nu\mu;ik}^*.
$
So,
\begin{equation}
    \mathrm{Re}(R_{\nu\mu;ik})=\mathrm{Re}(R_{\nu\mu;ki}),\quad \mathrm{Im}(R_{\nu\mu;ik})=-\mathrm{Im}(R_{\nu\mu;ki}).
    \label{eq:R-exchange-i-k}
\end{equation}
We can derive the following expression using a similar method,
\begin{equation}
    \mathrm{Re}(R_{\nu\mu;ik})=\mathrm{Re}(R_{\mu\nu;ik}),\quad \mathrm{Im}(R_{\nu\mu;ik})=-\mathrm{Im}(R_{\mu\nu;ik}).
    \label{eq:R-exchange-mu-nu}
\end{equation}
From Eq. (\ref{eq:H-exchange-mu-nu}) and Eq. (\ref{eq:R-exchange-mu-nu}), one can easily infer that the real part of the new MSQGT is symmetric while the imaginary part is antisymmetric under the exchange of $\nu$ and $\mu$. Besides, based on Eq. (\ref{eq:R-exchange-i-k}), one can derive that
\begin{equation}
\begin{aligned}
&\sum_{i,k=0}^{N-1}\frac{4p_i^2p_k}{(p_i+p_k)^2}\mathrm{Re}\langle  \partial_{\nu}\xi_{i}|\xi_{k}\rangle  \langle  \xi_{k}|\partial_{\mu}\xi_{i}\rangle\\
&=\sum_{i,k=0}^{N-1}\frac{4p_i^2p_k}{(p_i+p_k)^2}\mathrm{Re}(R_{\nu\mu;ik})\\
&=\sum_{i,k=0}^{N-1}\frac{2p_i^2p_k}{(p_i+p_k)^2}\mathrm{Re}(R_{\nu\mu;ik})+\sum_{i,k=0}^{N-1}\frac{2p_i^2p_k}{(p_i+p_k)^2}\mathrm{Re}(R_{\nu\mu;ik})\\
&=\sum_{i,k=0}^{N-1}\frac{2p_i^2p_k}{(p_i+p_k)^2}\mathrm{Re}(R_{\nu\mu;ik})+\sum_{i,k=0}^{N-1}\frac{2p_k^2p_i}{(p_k+p_i)^2}\mathrm{Re}(R_{\nu\mu;ki})\\
&=\sum_{i,k=0}^{N-1}\frac{2p_i^2p_k}{(p_i+p_k)^2}\mathrm{Re}(R_{\nu\mu;ik})+\sum_{i,k=0}^{N-1}\frac{2p_k^2p_i}{(p_k+p_i)^2}\mathrm{Re}(R_{\nu\mu;ik})\\
&=\sum_{i,k=0}^{N-1}\frac{2p_ip_k}{(p_i+p_k)}\mathrm{Re}(R_{\nu\mu;ik}).\\
\end{aligned}
\end{equation}
Substituting this equation into the real part of the MSQGT in Eq. (\ref{eq:QGT-expan}) yields the Eq. (\ref{eq:qgt-real}) presented in the main text.

\section{Derivation of the Phase-Arbitrariness-Independent Expression for the MSQGT}
\label{appendix:phase-arbitrariness-independent-qgt-appendix}
In this appendix, we provide the proof of Eq. \eqref{eq:QGT-calculate}.
First, we consider the representation of the derivative of density matrix $\rho=\sum_{i=0}^{N-1}p_i|\xi_i\rangle\langle \xi_i|$ in its eigenbasis, i.e.,
\begin{equation}
\begin{aligned}
&\langle  \xi_{i}|\partial_{\nu}\rho|\xi_{k}\rangle\\
&=\langle  \xi_{i}|\left(\sum_{l=0}^{N-1}\partial_{\nu}p_{l}|\xi_{l}\rangle  \langle  \xi_{l}|+p_{l}|\partial_{\nu}\xi_{l}\rangle  \langle  \xi_{l}|+p_{l}|\xi_{l}\rangle  \langle  \partial_{\nu}\xi_{l}|\right)|\xi_{k}\rangle\\
&=\partial_{\nu}p_{i}\delta_{ik}+p_{k}\langle  \xi_{i}|\partial_{\nu}\xi_{k}\rangle  +p_{i}\langle  \partial_{\nu}\xi_{i}|\xi_{k}\rangle\\
&=\partial_{\nu}p_{i}\delta_{ik}+(p_{k}-p_i)\langle  \xi_{i}|\partial_{\nu}\xi_{k}\rangle.
\end{aligned}
\end{equation}
where we have used the property $\partial_\nu\langle \xi_i|\xi_j\rangle=\langle \partial_\nu \xi_i|\xi_j\rangle+\langle \xi_i|\partial_\nu \xi_j\rangle=\partial_\nu \delta_{ij}=0$.
Then, we have 
\begin{equation}
\begin{aligned}
&\langle  \xi_{i}|\partial_{\nu}\rho|\xi_{k}\rangle  \langle  \xi_{k}|\partial_{\mu}\rho|\xi_{i}\rangle\\
&=\partial_{\nu}p_{i}\partial_{\mu}p_{i}\delta_{ik}+(p_{k}-p_i)(p_i-p_k)\langle  \xi_{i}|\partial_{\nu}\xi_{k}\rangle \langle  \xi_{k}|\partial_\mu\xi_{i}\rangle\\
&=\partial_{\nu}p_{i}\partial_{\mu}p_{i}\delta_{ik}+(p_{i}^{2}+p_{k}^{2}-2p_{i}p_{k})\langle  \partial_{\nu}\xi_{i}|\xi_{k}\rangle  \langle  \xi_{k}|\partial_{\mu}\xi_{i}\rangle\\
&=\partial_{\nu}p_{i}\partial_{\mu}p_{i}\delta_{ik}+[(p_{i}+p_{k})^{2}-4p_{i}p_{k}]\langle  \partial_{\nu}\xi_{i}|\xi_{k}\rangle  \langle  \xi_{k}|\partial_{\mu}\xi_{i}\rangle,
\label{eq:derivation-01}
\end{aligned}
\end{equation}
Utilizing expression \eqref{eq:derivation-01}, one can further obtain
\begin{equation}
\begin{aligned}
&\sum_{i,k=0}^{N-1} \frac{p_{i}}{(p_{i}+p_{k})^{2}}\langle  \xi_{i}|\partial_{\nu}\rho|\xi_{k}\rangle  \langle  \xi_{k}|\partial_{\mu}\rho|\xi_{i}\rangle\\
&= \sum_{i=0}^{N-1} \frac{p_{i} \partial_{\nu}p_{i}\partial_{\mu}p_{i}}{4p_{i}^{2}}+\sum_{i,k=0}^{N-1}p_{i}\langle  \partial_{\nu}\xi_{i}|\xi_{k}\rangle  \langle  \xi_{k}|\partial_{\mu}\xi_{i}\rangle  \\
&\quad -\sum_{i,k=0}^{N-1} \frac{4p_{i}^{2}p_{k}}{(p_{i}+p_{k})^{2}}\langle  \partial_{\nu}\xi_{i}|\xi_{k}\rangle  \langle  \xi_{k}|\partial_{\mu}\xi_{i}\rangle  \\
&=\sum_{i=0}^{N-1} \partial_{\nu}\sqrt{p_{i}}\partial_{\mu}\sqrt{p_{i}}+\sum_{i=0}^{N-1}p_{i}\langle  \partial_{\nu}\xi_{i}|\partial_{\mu}\xi_{i}\rangle  \\
&\quad-\sum_{i,k=0}^{N-1} \frac{4p_{i}^{2}p_{k}}{(p_{i}+p_{k})^{2}}\langle  \partial_{\nu}\xi_{i}|\xi_{k}\rangle  \langle  \xi_{k}|\partial_{\mu}\xi_{i}\rangle. 
\end{aligned}   
\end{equation}
According to Eq.~\eqref{eq:QGT-expan}, we obtain
\begin{equation}
Q_{\nu \mu}=\sum_{i,k=0}^{N-1} \frac{p_{i}}{(p_{i}+p_{k})^{2}} \langle  \xi_{i}|\partial_{\nu}\rho|\xi_{k}\rangle  \langle  \xi_{k}|\partial_{\mu}\rho|\xi_{i}\rangle.
\label{eq:QGT-calculate-appendix}
\end{equation}

\section{Derivation of the MSQGT as Mixed States Approach Pure States}
\label{appendix:mix-to-pure}
For a full-rank density matrix $\rho = \sum_{i=1}^N p_{i}|\xi_{i}\rangle\langle \xi_{i}|$, according to Eq. \eqref{eq:QGT-calculate}, the expression for MSQGT is
\begin{equation}
Q_{\nu \mu} = \sum_{i,k=0}^{N-1} \frac{p_{i}}{(p_{i}+p_{k})^{2}}
\langle \xi_{i}|\partial_{\nu}\rho|\xi_{k}\rangle 
\langle \xi_{k}|\partial_{\mu}\rho|\xi_{i}\rangle.
\label{eq:Q_numu}
\end{equation}
Since this equation is derived based on a full-rank density matrix, we cannot directly take $\rho$ as a pure state. However, we can use the following method to make it approach a pure state. Let us assume the eigenvalues of $\rho$ is given by
\begin{equation}
p_{i}=\frac{e^{-\beta E_{i}}}{Z},
\end{equation}
where $Z=\sum_{i=0}^{N-1} e^{-\beta E_i}$ and $E_i$ satisfying $0 \equiv E_{0} < E_{1} \leq E_{2} \leq E_{3} \leq \cdots \leq E_{N}$. When $\beta\to +\infty$, the mixed state reduce to a pure state $|\xi_{0}\rangle\langle\xi_{0}|$ and $Z\to 1$.

Let us examine some basic quantities that will be used in the following derivation under the limit $\beta\to +\infty$. First, for $\partial_\nu p_i$, it can be calculated as  
\begin{equation}
\begin{aligned}
\partial_\nu p_i
&=\frac{-Z\beta\partial_\nu E_i e^{-\beta E_i}-\partial_\nu Z e^{-\beta E_i}}{Z^2}\\
&=\frac{-Z\beta\partial_\nu E_i e^{-\beta E_i}-\sum_{k\neq 0} (-\beta \partial_\nu E_k)e^{-\beta E_k} e^{-\beta E_i}}{Z^2}\\
&=\frac{-\beta e^{-\beta E_i}(Z\partial_\nu E_i -\sum_{k\neq 0} \partial_\nu E_k e^{-\beta E_k})}{Z^2}
\end{aligned}
\end{equation}
When $i=0$,  $\partial_\nu E_i=0$, then
\begin{equation}
    \lim_{\beta\to +\infty}\partial_\nu p_0=0.
    \label{eq:partial_nu_p_0}
\end{equation}

When $i\neq 0$,
\begin{equation}
\lim_{\beta \to +\infty}\partial_\nu p_i \propto \lim_{\beta \to +\infty} \beta e^{-\beta E_i} \partial_\nu E_i\to 0.
\label{eq:partial_nu_p_i}
\end{equation}
Since we are considering parameters that vary continuously, the derivative $\partial_{\nu}E_{i}$ is finite, resulting in $\lim_{\beta \to +\infty} \partial_\nu p_i\to 0$.

Based on the properties of $\partial_\nu p_i$ under the limit, we can derive that 
\begin{equation}
\begin{aligned}
&\lim_{\beta\to +\infty} \partial_\nu \rho\\
&=\lim_{\beta\to +\infty}\sum_i (\partial_\nu p_i|\xi_i\rangle\langle \xi_i|+p_i|\partial_\nu\xi_i\rangle\langle \xi_i|+p_i|\xi_i\rangle\langle \partial_\nu \xi_i|)\\
&=|\partial_\nu \xi_0\rangle\langle\xi_0|+|\xi_0\rangle\langle\partial_\nu \xi_0|
\label{eq:partia_nu_rho}.
\end{aligned}
\end{equation}

Based on these discussions, we can then calculate the MSQGT in the limit $\beta\to+\infty$. To simplify expressions, we denote the terms in Eq. \eqref{eq:Q_numu} by $q_{ik}$, i.e.,
\begin{equation}
q_{ik} = \frac{p_{i}}{(p_{i}+p_{k})^2}\langle \xi_{i}|\partial_{\nu}\rho|\xi_{k}\rangle 
\langle \xi_{k}|\partial_{\mu}\rho|\xi_{i}\rangle,
\end{equation}
Therefore, $Q_{\nu\mu}=\sum_{i,k=0}^{N-1}q_{ik}$. 
We can divide these terms into four categories:

1. $i=k=0$:
\begin{equation}
\begin{aligned}
\lim_{\beta\to+\infty}q_{00}
&= \lim_{\beta\to+\infty}\frac{p_{0}}{(p_{0}+p_{0})^{2}} \langle  \xi_{0}|\partial_{\nu}\rho|\xi_{0}\rangle  \langle  \xi_{0}|\partial_{\mu}\rho|\xi_{0}\rangle\\
&= \frac{1}{4}\lim_{\beta\to+\infty} (\partial_\nu p_0+\langle \xi_0|\partial_\nu \xi_0\rangle+\langle\partial_\nu \xi_0| \xi_0\rangle)\\
&\qquad\qquad \times (\partial_\mu p_0+\langle \xi_0|\partial_\mu \xi_0\rangle+\langle \partial_\mu\xi_0| \xi_0\rangle)\\
&=0,
\end{aligned}
\end{equation}
where we have used Eq. \eqref{eq:partial_nu_p_0} and the property $\partial_\nu\langle\xi_0|\xi_0\rangle=\langle  \xi_{0}|\partial_{\nu}\xi_{0}\rangle  +\langle  \partial_{\nu}\xi_{0}|\xi_{0}\rangle=0$.
  
2. $i=0,k\neq 0$:

Since $\lim_{\beta\to+\infty}q_{00}=0$, it follows that
  \begin{equation}
    \begin{aligned}
    &\lim_{\beta\to +\infty}\sum_{k\neq 0}^{N-1}q_{0k}\\
    &=\lim_{\beta\to +\infty}\sum_{k\neq 0}^{N-1} \langle  \xi_{0}|\partial_{\nu}\rho|\xi_{k}\rangle  \langle  \xi_{k}|\partial_{\mu}\rho|\xi_{0}\rangle  +4\lim_{\beta\to+\infty}q_{00}\\
    &=\lim_{\beta\to +\infty}\sum_{k=0}^{N-1} \langle  \xi_{0}|\partial_{\nu}\rho|\xi_{k}\rangle  \langle  \xi_{k}|\partial_{\mu}\rho|\xi_{0}\rangle  \\
    &=\lim_{\beta\to +\infty}\langle \xi_{0}|\partial_{\nu}\rho\partial_{\mu}\rho|\xi_{0}\rangle\\
    \end{aligned} 
    \end{equation}
Substituting Eq. \eqref{eq:partia_nu_rho} into this equation, we obtain
\begin{equation}
\lim_{\beta\to +\infty}\sum_{k\neq 0}^{N-1}q_{0k}=\langle  \partial_{\nu}\xi_{0}|\partial_{\mu}\xi_{0}\rangle  -\langle \partial_{\nu} \xi_{0}|\xi_{0}\rangle  \langle  \xi_{0}|\partial_{\mu}\xi_{0}\rangle.
\end{equation}
which is exactly the pure-state QGT.

3. $i\neq 0,k=0$:

\begin{equation}
\lim_{\beta\to +\infty} q_{i0}=0.
\end{equation}

4. $i\neq 0,k\neq 0$:
\begin{equation}
\begin{aligned}
q_{ik} &= \frac{p_{i}}{(p_{i}+p_{k})^2}\langle \xi_{i}|\partial_{\nu}\rho|\xi_{k}\rangle 
\langle \xi_{k}|\partial_{\mu}\rho|\xi_{i}\rangle \\
&=\frac{p_{i}}{(p_{i}+p_{k})^2}\Bigl(\partial_{\nu}p_{i}\,\delta_{ik}
+p_{k}\langle \xi_{i}|\partial_{\nu}\xi_{k}\rangle 
+p_{i}\langle \partial_{\nu}\xi_{i}|\xi_{k}\rangle\Bigr)\\
&\quad \times \Bigl(\partial_{\mu}p_{i}\,\delta_{ik}
+p_{i}\langle \xi_{k}|\partial_{\mu}\xi_{i}\rangle 
+p_{k}\langle \partial_{\mu}\xi_{k}|\xi_{i}\rangle\Bigr).
\label{eq:q_ik}
\end{aligned}
\end{equation}
To calculate the $q_{ik}$, we can first examine the behavior of the factors shown in the expansion of $q_{ik}$ in the limit $\beta\to+\infty$. 
Firstly, 
\begin{equation}
\begin{aligned}
&\lim_{ \beta \to +\infty } \frac{p_{i}^2p_{k}}{(p_{i}+p_{k})^2}\\
&=\lim_{ \beta \to +\infty } \frac{1}{Z}\frac{e^{-\beta(2E_{i}+E_{k})}}{e^{-2\beta E_{i}}+2e^{-\beta(E_{i}+E_{k})}+e^{-2\beta E_{k}}}\\
&=\lim_{ \beta \to +\infty } \frac{1}{e^{\beta E_{k}}+2e^{\beta E_{i}}+e^{\beta(2E_{i}-E_{k})}}\\
&\to 0.
\end{aligned}
\end{equation}
Similarly, the following factors can be obtain:
\begin{equation}
\lim_{ \beta \to +\infty } \frac{p_{i}p_{k}^2}{(p_{i}+p_{k})^2}\to 0,\quad 
\lim_{ \beta \to +\infty } \frac{p_{i}^3}{(p_{i}+p_{k})^2}\to 0.
\end{equation}
Next, we examine another type of factors, according to Eq. \eqref{eq:partial_nu_p_i}:
\begin{equation}
\begin{aligned}
&\lim_{ \beta \to +\infty } \frac{p_{i}p_{k}\partial_{\nu}p_{i}}{(p_{i}+p_{k})^2}\\
&\propto \lim_{ \beta \to +\infty } \frac{e^{-\beta E_i}e^{-\beta E_k}\beta e^{-\beta E_{i}}\partial_\nu E_i}
{e^{-2\beta E_{i}}+2e^{-\beta(E_{i}+E_{k})}+e^{-2\beta E_{k}}}\\
&=\lim_{ \beta \to +\infty } \frac{\beta|\partial_{\nu} E_{i}|}
{e^{\beta E_{k}}+2e^{\beta E_{i}}+e^{\beta(2E_{i}-E_{k})}}\\
&\to 0.
\end{aligned}
\end{equation}
As a result, the above factors vanish. By a similar analysis, one can show that
\begin{equation}
\begin{aligned}
&\lim_{ \beta \to +\infty } \frac{p_{i}^2 \partial_{\nu}p_{i}}{(p_{i}+p_{k})^2}
\to 0\\ 
&\lim_{ \beta \to +\infty } \frac{p_{i}^2\,\partial_{\mu}p_{i}}{(p_{i}+p_{k})^2}\to 0,\quad \\
&\lim_{ \beta \to +\infty } \frac{p_{i}p_{k}\,\partial_{\mu}p_{k}}{(p_{i}+p_{k})^2}\to 0.
\end{aligned}
\end{equation}
Furthermore, we have
\begin{equation}
\begin{aligned}
&\lim_{ \beta \to +\infty } \frac{p_{i}\,\partial_{\mu}p_{i}\,\partial_{\nu}p_{i}}{(p_{i}+p_{k})^2}\\
&\propto\lim_{ \beta \to +\infty } Z\frac{\beta^2(\partial_{\mu}E_{i}\,\partial_{\nu}E_{i})e^{-3\beta E_{i}}}
{e^{-2\beta E_{i}}+2e^{-\beta(E_{i}+E_{k})}+e^{-2\beta E_{k}}}\\
&=\lim_{ \beta \to +\infty } \frac{\beta^2(\partial_{\mu}E_{i}\,\partial_{\mu}E_{i})}{e^{\beta E_{i}}+2e^{2\beta (E_{i}-E_{k})}+e^{\beta(3E_{i}-2E_{k})}}\\
&\to 0.
\end{aligned}
\end{equation}
Substituting these factors into Eq. \eqref{eq:q_ik}, we can deduce that
\begin{equation}
\lim_{ \beta \to +\infty } q_{ik}=0,\quad \text{for } i \neq 0, k \neq 0.
\end{equation}

To sum up, when $\rho\to |\xi_0\rangle\langle \xi_0|$, 
\begin{equation}
Q_{\nu \mu}\to\langle  \partial_{\nu}\xi_{0}|\partial_{\mu}\xi_{0}\rangle  -\langle  \partial_\nu\xi_{0}|\xi_{0}\rangle  \langle  \xi_{0}|\partial_{\mu}\xi_{0}\rangle.
\end{equation}

\section{Derivation of the Geodesic Equation}
\label{appendix:geodesic}
In this Appendix, we utilize the variational method to derive the geodesic equation.
Let us consider an curve $|\psi'(t)\rangle$ which is very close to $|\psi(t)\rangle$ and $|\psi'(t)\rangle=|\psi(t)\rangle +|\delta\psi(t)\rangle$. Then
\begin{equation}
\langle  \psi'|\psi'\rangle  \approx \langle  \psi|\psi \rangle  +\langle  \psi|\delta \psi \rangle  + \langle  \delta \psi|\psi \rangle=1.
\end{equation}
To preserve the normalization of $|\psi'\rangle$, we must require
\begin{equation}
\langle  \psi|\delta \psi \rangle +\langle  \delta \psi|\psi \rangle  =2(|\psi \rangle  ,|\delta \psi \rangle  )=0.
\label{eq:variaiton-condition}
\end{equation}
where $(\cdot,\cdot)$ denotes the inner product defined in Eq. (\ref{eq:inner-product-def}). Note that this inner product is linear only under real scalar multiplication. Specifically, when $a$ is a real number, we have $(a|A\rangle ,|B\rangle )=a(|A\rangle ,|B\rangle )$, whereas for complex $b$, it follows that $(b|A\rangle ,|B\rangle )=(b^*\langle A|B\rangle +b\langle B|A\rangle)/2\neq b(|A\rangle ,|B\rangle)$.

For covariant derivative, we have
\begin{equation}
\begin{aligned}
\delta|D_{t}\psi \rangle
&=\delta(|\partial_{t}\psi \rangle  -i\mathcal{A}_{t}|\psi \rangle )\\
&=|\partial_{t}\delta \psi \rangle  -i\delta \mathcal{A}_{t}|\psi \rangle  -i\mathcal{A}_{t}|\delta \psi \rangle\\
&=|D_{t} \delta \psi \rangle  -i \delta \mathcal{A}_{t}|\psi \rangle  
\end{aligned},
\end{equation}
where $\mathcal{A}_{t}$ and $\mathcal{A}'_{t}$ are connections for $|\psi(t)\rangle$ and $|\psi'(t)\rangle$ respectively. $\delta \mathcal{A}_{t}=\mathcal{A}'_{t}-\mathcal{A}_{t}$ is in general not zero. Thus $\delta$ and $D_{t}$ do not commute. However, both connection $\mathcal{A}'_{t}$ and $\mathcal{A}_{t}$ are Hermitian, it follows that $\delta \mathcal{A}_{t}$ is Hermitian. Consequently, $i\delta \mathcal{A}_{t}|\psi \rangle$ is a vertical vector according to Eq. \eqref{eq:vertical-vector} and the inner product between it and the horizontal vector $|D_{t}\psi \rangle$ should be 0, i.e.
\begin{equation}
(|D_{t}\psi \rangle ,i\delta \mathcal{A}_{t}|\psi \rangle )=0.
\end{equation}
Substitute this condition into Eq. (\ref{eq:variation of line length}), we can obtain
\begin{equation}
\begin{aligned}
\delta l&=\int_{0}^{T} (|D_{t}\psi \rangle  ,|D_{t}\delta \psi \rangle  ) dt\\
&=\frac{1}{2}\int_{0}^{T} (\langle  D_{t}\psi|D_{t}\delta \psi \rangle+\langle  D_{t}\delta \psi|D_{t}\psi \rangle  )dt \\
&=\frac{1}{2}\int_{0}^{T} [\langle  D_{t}\psi|\partial_{t}\delta \psi \rangle  -i\langle  D_{t}\psi|\mathcal{A}_{t}|\delta\psi \rangle\\
&\qquad\qquad+\langle  \partial_{t}\delta \psi|D_{t}\psi \rangle  +i\langle  \delta\psi|\mathcal{A}_{t}|D_{t}\psi \rangle  ]dt.
\end{aligned}
\end{equation}
Since we aim to find the geodesic between two points, we must keep the endpoints fixed during variation, i.e.,$|\delta\psi(0)\rangle=|\delta\psi(T)\rangle=0$. Applying integration by parts and utilizing this condition, we obtain
\begin{equation}
\begin{aligned}
\delta l
&=\frac{1}{2}\int_{0}^{T} [-(\partial_{t}\langle D_{t}\psi|)|\delta \psi \rangle ]-i\langle  D_{t}\psi|\mathcal{A}_{t}|\delta \psi \rangle \\
&\qquad\qquad-\langle  \delta \psi|(\partial_{t}|D_{t}\psi \rangle) +i\langle  \delta \psi|\mathcal{A}_{t}|D_{t}\psi \rangle  ]dt\\
&=-\frac{1}{2}\int_{0}^{T}[(\partial_{t}\langle  D_{t}\psi|+i\langle  D_{t}\psi|\mathcal{A}_t)|\delta \psi \rangle  \\
&\qquad\qquad+\langle  \delta \psi|(\partial_{t}|D_{t}\psi \rangle  -i\mathcal{A}_{t}|D_{t}\psi \rangle  )]dt \\
&=-\frac{1}{2}\int_{0}^{T}[\langle  D_{t}D_{t}\psi|\delta \psi \rangle +\langle  \delta \psi|D_{t}D_{t}\psi \rangle]  dt\\
&=-\int_{0}^{T}(|D_{t}D_{t}\psi \rangle  ,|\delta \psi \rangle  )dt.
\end{aligned}
\end{equation}
To ensure that $\delta l=0$ holds for arbitrary $|\delta \psi \rangle$, combining the condition (\ref{eq:variaiton-condition}) with the real linearity of the real inner product, we derive the extremal condition for the path length as:
\begin{equation}
|D_{t}D_{t}\psi \rangle  =C(t)|\psi \rangle.
\end{equation}
where $C(t)$ is a real function.

Substituting this condition into the properties of the second covariant derivative in Eq.(\ref{eq:second_covariant_derivative}), we obtain
$C(t)=-\langle D_{t}\psi|D_{t}\psi \rangle$. Hence, the geodesic equation is
\begin{equation}
|D_{t}D_{t}\psi \rangle=-\langle  D_{t}\psi|D_{t}\psi \rangle  |\psi \rangle=-|\psi\rangle,
\end{equation}
where we have used the condition in Eq. (\ref{eq:dl-rescale}).

\end{document}